\documentclass[aps,onecolumn,preprint,superscriptaddress,nofootinbib,prd]{revtex4}
\usepackage{amsmath,amssymb,color,mathrsfs, graphicx,verbatim,epsfig, bbm, wasysym}

\def\lsim{\mathrel{\rlap{\lower4pt\hbox{\hskip1pt$\sim$}}
    \raise1pt\hbox{$<$}}}
\def\gsim{\mathrel{\rlap{\lower4pt\hbox{\hskip1pt$\sim$}}
    \raise1pt\hbox{$>$}}}

\newcommand{\be}{\begin{eqnarray}}
\newcommand{\ee}{\end{eqnarray}}

\def\addresses#1#2{\hbox to \hsize{\@tablebox{#1}\hfil\@tablebox{#2}}}
\def\@tablebox#1{\vtop{\hsize=5in \begin{flushleft} #1 \end{flushleft}}}

\def\beq{\begin{equation}}
\def\eeq{\end{equation}}
\def\bit{\begin{itemize}}
\def\eit{\end{itemize}}
\def\cL{{\cal L}}
\def\beqarray{\begin{eqnarray}}
\def\eeqarray{\end{eqnarray}}

\def\bbbar{$b\bar{b}$}

\def\WpWm{$W^+W^-$}

\def\mathmet{\displaystyle{\not}E_T}

\begin{document}

\begin{titlepage}

\thispagestyle{empty}

\begin{flushright}UMD-PP-10-017
\end{flushright}
\vspace{0.2cm}

\begin{center}

\vskip .5cm

{\Large \bf Jet Substructure and the Search for Neutral Spin-One\\ \vspace{0.3cm} 
Resonances in Electroweak Boson Channels}
\vskip 1.0cm
{\large Andrey Katz,$^1$ Minho Son,$^2$ and Brock Tweedie$^{3,4}$}
\vskip 0.4cm
{\it $^1$Department of Physics, University of Maryland, College Park, MD 20742} \\
{\it $^2$Department of Physics, Yale University, New Haven, CT 06511} \\
{\it $^3$Department of Physics and Astronomy, Johns Hopkins University, Baltimore, MD 21218} \\
{\it $^4$Physics Department, Boston University, Boston, MA 02215}
\vskip 1.2cm

\end{center}

\noindent
Strongly coupled models at the TeV scale often predict one or more neutral spin-one resonances ($Z'$) which have appreciable branching fractions to electroweak bosons, namely the Higgs and longitudinal $W$ and $Z$.  These resonances are usually believed to have multi-TeV mass due to electroweak precision constraints, placing them on the edge of LHC discovery reach.  Searching for them is made particularly challenging because hadronically decaying electroweak bosons produced at such high energy will appear very similar to QCD jets.  In this work we revisit the possibility of discovering these resonances at the LHC, taking advantage of recently developed jet substructure techniques.  We make a systematic investigation of substructure performance for the identification of highly Lorentz-boosted electroweak bosons, which should also be applicable to more general new physics searches.  We then estimate the model-independent $Z'$ discovery reach for the most promising final-state channels, and find significant improvements compared to previous analyses.  For modes involving the Higgs, we focus on a light Higgs decaying to $b\bar b$.  We further highlight several other novelties of these searches.  In the case that vertex-based $b$-tagging becomes inefficient at high $p_T$, we explore the utility of a muon-based $b$-tag, or no $b$-tag at all.  We also introduce the mode $Z'\to Zh \to (\nu\bar\nu)(b\bar b)$ as a competitive discovery channel.

\end{titlepage}

%\newpage 
\setcounter{page}{1}

%%%%%%%%%%%%%%%%%%%%%%%%%%%%%%%%%%%%%%%%%%%%%%%%%%%%%%%%%%%

%\tableofcontents

\section{Introduction}
\label{sec:intro}

%%%%%%%%%%%%%%%%
% Introduction
%%%%%%%%%%%%%%%%

In the next few years, expectations are high that the LHC will discover the Higgs boson, finally verifying all of the basic building blocks of the Standard Model.  There is also much anticipation that this discovery will not signal the end of novel physics near the electroweak scale.  However, given the lack of clear anomalies in existing flavor and electroweak precision measurements, a very real possibility is that the Standard Model with a light elementary Higgs will turn out to be a reasonably good description of LHC phenomenology at energies of less than about a TeV.  Should this prove to be the case, after a few 10's of fb$^{-1}$ of data attention will become highly focused on carefully testing the properties of the Higgs through a combination of coupling measurements~\cite{Rainwater:2007cp} and investigations of $W$, $Z$, and $h$ production at progressively higher energies~\cite{Bagger:1995mk,Butterworth:2002tt}.  One of the most important goals is to determine to what extent the Higgs is truly an elementary particle, versus a light state arising from a new composite sector.  

Like most scenarios for physics beyond the Standard Model, specific composite Higgs models are under a certain amount of tension between, on the one hand, satisfying the flavor and electroweak precision constraints and, on the other hand, reintroducing the fine-tuning which the models were invented to solve, though still at a much reduced level (see for reviews e.g.~\cite{Schmaltz:2005ky,Davoudiasl:2009cd}).  The usual compromise is to choose a compositeness scale of several TeV.  If this kind of scenario proves to be true, then one of the main hints that there is something unusual about the Higgs will be the production of multi-TeV resonances with large couplings to Higgs bosons.

In this paper, we aim to make a detailed investigation of the LHC's sensitivity to such scenarios via the production of a neutral spin-1 resonance in $q\bar{q}$ annihilation.  The existence of one or more $Z'$ particles with these quantum numbers is rather ubiquitous in composite Higgs models, for example serving as the analogs of the $\rho$ mesons of QCD or as higher states of an enlarged broken gauge sector.  Owing to the large coupling with the Higgs, as well as with the (longitudinal) $W$ and $Z$ bosons, the decays of these $Z'$ resonances will typically be dominated by the modes $Zh$ and \WpWm.  This is in stark contrast to more standard $U(1)'$ models, which usually predict much larger relative rates into leptons and quarks.  Observation of a $Z'$ with large branching fraction to either of these electroweak boson modes would therefore serve as highly suggestive evidence for a composite Higgs sector, whereas lack of an observation would place important constraints.

The two intermediate states $Zh$ and \WpWm\ can lead to a variety of final-state configurations.  Despite the additional structure compared to, say, a simple dileptonic final state, these can be some of the most challenging search channels.  Indeed, the flavor and electroweak data hint at the most difficult possible situation:  the Higgs may be light and the $Z'$ may be heavy.  The former presents a difficulty since the Higgs decays then become dominated by $b\bar b$, versus the more distinctive $WW^{(*)}/ZZ^{(*)}$ modes of heavier Higgses.  The latter suggests that $Z'$ production rates will be small (e.g., $O$(1 fb) for $m_{Z'} = 3$ TeV~\cite{Agashe:2007ki,Han:2003wu}), and also that the secondary decay products of the highly energetic electroweak bosons will be Lorentz-boosted into tight, jet-like configurations in the detector.  For leptonic decays of the $W$ and $Z$ bosons, this last observation does not necessarily pose a problem, beyond the usual loss of efficiency due to leptonic branching fraction.  But for any boson decays involving hadrons, the distinction with a QCD-induced jet can become uncomfortably blurred.

In the past several years, various techniques have been developed for dealing with highly boosted $W$s, $Z$s, Higgses, and tops (e.g.,~\cite{Seymour:1993mx,Butterworth:2002tt,Butterworth:2008iy,Brooijmans:2008zza,Thaler:2008ju,Kaplan:2008ie,Ellis:2009su,Butterworth:2009qa,Kribs:2010hp,Plehn:2009rk,Plehn:2010st,Rehermann:2010vq,Almeida:2008yp,Almeida:2010pa}.  For our present study with boosted electroweak bosons, we will utilize the procedure of~\cite{Butterworth:2008iy}.  Originally, this was developed to aid in the identification of semi-boosted light Higgses in the high-$p_T$ tail of Standard Model Higgsstrahlung production.  The procedure adapts on an event-by-event basis to pick out hard substructures within a region of interest in the detector, while actively attempting to remove uncorrelated radiation which can degrade mass resolution.  We will find that the ideas of~\cite{Butterworth:2008iy} easily extrapolate into the truly high-boost regime for the Higgs, as well as to hadronic decays of boosted $W$s and $Z$s.  Along the way, we will find that the scale-invariant nature of the procedure offers a number of nontrivial advantages.

Beyond the simple \bbbar\ decays of light Higgses, as well as $q\bar{q}$ decays of electroweak gauge bosons, substructure-type techniques can also be applied to the other possible decay modes of a boosted Higgs, such as $\tau^+\tau^-$ and $WW^{(*)}/ZZ^{(*)}$.  Including these modes in a $Z'\to Zh$ search offers several advantages.  First, a discovery in multiple modes would bolster confidence that we are seeing a real signal.  Second, broadening the set of Higgs decay modes also broadens the sensitivity over Higgs masses, and allows for increased sensitivity in cases with nonstandard Higgs couplings (e.g., gaugephobic~\cite{Cacciapaglia:2006mz,Galloway:2008yh,Galloway:2009xn} or fermiophobic~\cite{Diaz:1994pk}).  Third, a measurement of Higgs branching fraction ratios gives us further clues regarding its identity (especially whether it is the same ``Higgs'' as the one discovered at lower energies) and, in the Standard Model limit, can give us a secondary handle on the mass.  In this paper, we focus on the specific case of a 120 GeV Higgs decaying to $b\bar b$.  We separately explore the decays $h\to\tau^+\tau^-$ and $h\to WW^{(*)}/ZZ^{(*)}$ in two companion papers~\cite{ZprimesTau,Zprimes2}.  (See also~\cite{Barger:2009xg} for an analysis of $Z'\to Zh\to Z(ZZ)$ in the all-leptonic channel.)

Given this restriction, we find that the most promising $Z'$ discovery channels are 
\bit
\item $WW \to (l\nu)(q\bar q')$
\item $Zh \to (l^+l^-)(b\bar b)$
\item $Zh \to (\nu\bar\nu)(b\bar b)$
\eit
The first two of these have been explored before in~\cite{Benchekroun:2001je,Atl:LH,Agashe:2007ki}.  (The first channel was also studied in~\cite{Butterworth:2002tt,Aad:2009wy} in the context of weak boson fusion.  We also point out a study in~\cite{Hackstein:2010wk}, closely related to our second channel, of $h\to ZZ\to(l^+l^-)(q\bar q)$ at $m_{h}=$ 300-600 GeV using substructure methods.)  By applying jet substructure techniques, we obtain significant improvements, for example achieving sensitivity to fb-scale cross sections for $m_{Z'} = 3$ TeV in \WpWm\ with only 100~fb$^{-1}$ of data from a 14 TeV LHC.  The last mode, utilizing the invisible $Z$ decay in $Zh$, appears to have been largely underappreciated for resonance searches, but we find that it is quite competitive with the leptonic $Z$.  We further explore the utility of $b$-tagging the Higgs-jet in the two $Zh$ modes, using a muon-based technique that should extrapolate robustly to high-$p_T$.  Application of such a tag appears to be beneficial for finding $Z'$ bosons in roughly the 1-2 TeV mass range for the leptonic $Z$ mode, and up to about 3 TeV for the invisible $Z$.  Beyond these masses, the background becomes so small that an untagged analysis is more sensitive.

The paper is organized as follows.  In section~\ref{sec:theory} we take a more detailed look at the theoretical motivation for our study and classify the relevant $Z'$ final states.  In section~\ref{sec:substructure} we review jet substructure techniques, characterize their performance for tagging boosted electroweak bosons, and make a comparison with a more traditional jet-based approach.  We present our model-independent discovery reach analyses in section~\ref{sec:WW}.  Section~\ref{sec:conclusion} contains conclusions.  We include discussions of pileup and detector modeling in appendices~\ref{sec:pileup} and~\ref{sec:detector}, respectively.

\section{$Z'$ Theory and Final-State Channels}
\label{sec:theory}

%%%%%%%%%%%%%%%%%%%%%
% Section on theory
%%%%%%%%%%%%%%%%%%%%%

\subsection{Theoretical motivation}

For our purposes, the relevant couplings of a lone generic $Z'$ can be parametrized as\footnote{We neglect four-point couplings with the SM Higgs, as well as terms induced by replacing $\partial^\mu\to D^\mu$.  (These are both important for a more complete, gauge-invariant treatment of $Z'$ couplings to the SM gauge bosons.)  The possibility of kinetic mixing with hypercharge is already incorporated in our parametrization.  The field redefinitions necessary to undo such a mixing will simply shift ${j'}^\mu \to {j'}^\mu + \kappa\, j_Y^\mu$, where $\kappa$ is a (usually small) number proportional to the coefficient of the mixing term.}
\beq\label{paremetrization}
\cL_{\rm int} =  {Z'}_\mu {j'}^\mu \ \ \ {\rm with } \ \ \ {j'}^\mu = \sum_i g_i' \psi_i^\dagger \bar\sigma^\mu \psi_i + g_H'\, (H^\dagger i\partial^\mu H + h.c.).
\eeq   
Here, $\psi_i$ stands for the chiral Standard Model (SM) fermions (in complete $SU(2)_L$ multiplets) and $H$ for the Higgs doublet (neutral Higgs will be denoted by $h$).  $g_i'$ and $g_H'$ are the respective coupling constants.  The coupling $g_H$ controls the decay rate of the $Z'$ into $Zh$ and \WpWm\ by virtue of the Goldstone boson equivalence theorem.  (We exclusively work in the limit $m_{Z'} \gg m_Z,m_W$.)  Immediate consequences of this are that the rates into $Zh$ and \WpWm\ are nearly equal, and the $Z$s and $W$s produced in the decays are highly longitudinally polarized.  More general models (as we will see below) can have more than one $Z'$ which can mix with each other and with the $Z$, leading to $SU(2)_L$-nonuniversalities in the couplings of the individual $Z'$ mass eigenstates.  However, the approximate equality of total $Zh$ and \WpWm\ rates is usually maintained, as electroweak symmetry breaking (EWSB) is necessarily a small effect in the heavy $Z'$ sector.

Roughly speaking, $Z'$ models can be categorized into two classes.  In the first class, the $Z'$ is realized as a gauge boson of a new gauge group broken near the TeV scale.  (See~\cite{Langacker:2008yv} for a review.)  In the second class, the $Z'$ emerges from a new strongly-coupled sector which also generates the Higgs boson or directly breaks electroweak symmetry.  One simple example of the former class is an admixture of $U(1)_Y$ and $U(1)_{B-L}$.  Another typical example is an admixture of the $U(1)_\chi$ and $U(1)_\psi$ subgroups of $E_6$.\footnote{The first example is automatically anomaly free without extra exotic fermions other than right-handed neutrinos.  The second example requires new exotic fermions near the TeV scale.}  The class of composite $Z'$ includes various models of Technicolor~\cite{Gildener:1976ih, Farhi:1979zx, Holdom:1984sk, Appelquist:1986an,Bando:1987we,Miransky:1984ef,Appelquist:1987fc,Csaki:2003zu,Luty:2004ye}, composite Higgs~\cite{Kaplan:1983fs,Kaplan:1983sm,Georgi:1984af,Dugan:1984hq,Contino:2003ve}, partial compositeness~\cite{Kaplan:1991dc}, and warped 5D models~\cite{Randall:1999vf,Davoudiasl:1999tf,Pomarol:1999ad,Agashe:2003zs}.  The Little Higgs~\cite{ArkaniHamed:2001nc,Schmaltz:2005ky} effectively lives in between these two classes.

The simple gauge $Z'$ couples fairly democratically between fermions and Higgs, usually leading to much higher branching fractions into fermions simply due to multiplicity (as well as a factor of two enhancement for decays into fermions versus decays into scalars).  In fact, when the lepton Yukawa couplings have SM-like minimal structure, and respect the $U(1)'$ symmetry, we can make a sharper statement:
\beq\label{BRratio}
BR(Z'\rightarrow l^+l^-) \geq BR(Z' \rightarrow Zh\ {\rm or}\ W^+W^-),
\eeq
where $l$ is any individual species of lepton.
If this relation is violated, we can infer that either the lepton Yukawa sector has non-minimal structure (such as multiple Higgs doublets or $U(1)'$-breaking spurions) or the couplings of the $Z'$ do not furnish a gauge symmetry.  Note that for a hypercharge-sequential $Z'$, the branching fraction to each species of lepton is 12\%, whereas the branching fraction to $Zh$ or \WpWm\ is $1.2$\%.  For a $B-L$ boson, the branching fraction to each lepton is 15\%, and into electroweak bosons it is zero.   It is important to realize that even if a resonance is first discovered in the $l^+l^-$ modes (which are very straightforward experimentally), concurrent searches in $Zh$ and \WpWm\ can yield very valuable information about the nature of the resonance, even in the case that nothing is found.

The situation drastically changes if the both the $Z'$ and the Higgs doublet are composite particles generated by a new strongly-coupled sector.  In this case, the coupling to the Higgs doublet can be hierarchically larger than the couplings to the fermions, and the decays of the $Z'$ become dominated by the modes $Zh$ and \WpWm\ (and perhaps $t\bar t$, as we will see).\footnote{It is possible in such composite scenarios that the Higgs boson proper, the neutral $CP$-even scalar component of the doublet, is not a well-defined particle.  This is the case in Technicolor and other Higgsless theories.  The decays of the $Z'$ may then become completely dominated by \WpWm\ (in close analogy with $\rho^0\to\pi^+\pi^-$ in QCD), or by other light composites.  As we ultimately perform a model-independent analysis in this paper, in particular an independent study of \WpWm, our results can also be applied to these cases if $m_{Z'} \gsim 1$ TeV.}  On the other hand, the typically small couplings between the composite sector and the fundamental SM fermions can significantly reduce the production cross section via $q\bar q$ annihilation.  Indeed, weak boson fusion is often taken as the most robust way to produce composite resonances, but a very high price in effective luminosity (see, e.g.,~\cite{Butterworth:2002tt}).  Nonetheless, realistic and well-motivated models have been identified where the competition between small production rate via $q\bar q$ and large decay rate into electroweak bosons can roughly compensate.  A very well-studied class of such models are the warped 5D constructions inspired by the Randall-Sundrum (RS) model~\cite{Randall:1999vf}, with the SM fields propagating in the bulk~\cite{Davoudiasl:1999tf,Pomarol:1999ad}.  According to AdS/CFT duality~\cite{Maldacena:1997re}, these can be interpreted as weakly-coupled holographic duals of large-$N_c$ 4D field theories at very strong coupling and which exhibit near-conformality above the TeV scale~\cite{ArkaniHamed:2000ds,Rattazzi:2000hs}.

As an illustrative example, we briefly review the minimal custodial RS model with bulk SM fields~\cite{Agashe:2003zs}.  (A more detailed review of the $Z'$ phenomenology of this model appears in~\cite{Agashe:2007ki}.)  Implementing a gauged custodial symmetry in the bulk (the holographic dual of a global custodial symmetry of the strongly-coupled sector) prevents large shifts of the electroweak $T$ parameter, and can be modified to better accommodate measurements of $Z\to b \bar b$~\cite{Agashe:2006at}.  The complete bulk gauge symmetry of the electroweak sector is $SU(2)_L\times SU(2)_R\times U(1)_{B-L}$, and boundary conditions are chosen so that only the $SU(2)_L\times U(1)_Y$ subgroup has zero modes (which subsequently acquire mass through the usual Higgs mechanism).  The higher KK excitations sample from the complete gauge group, with each KK level containing two $W'$ bosons and three $Z'$ bosons.  All of the bosons have nearly degenerate masses set by the 5D geometry, and can be highly mixed after EWSB.

In this scenario, the couplings of the $Z'$ bosons to the SM fermions and the Higgs doublet have a natural geometrical interpretation.  The couplings are determined by the overlaps among the wavefunctions along the fifth dimension.  In order to solve the hierarchy problem as in the original RS, the Higgs profile is localized near (or confined to) the IR brane.  Light fermions are localized near the UV brane, leading to small overlap with the Higgs, and therefore small 4D Yukawa couplings.  Heavy fermions such as the top quark are localized toward the IR brane.  The gauge zero modes are uniformly spread throughout the space, but their KK excitations are localized near the IR brane.  Consequently, the couplings of $Z'$ to light fermions are suppressed, whereas the couplings to the Higgs doublet (and top quarks) are enhanced.  As the region near the UV brane represents non-composite physics in the dual description, whereas the bulk and the IR brane represent composite physics, the tail of the KK wavefunctions near the UV brane indicates the kind of non-composite/composite mixing that occurs between the photon and $\rho^0$ in QCD.  This mixing can be significant enough to allow reasonable production rates in $q\bar q$ annihilation.

However, mixing between the composite sector and light SM sector also leads to tension, as there will be associated shifts in the values of electroweak and flavor observables.  In order to satisfy electroweak constraints, coming mainly from the $S$ parameter, the mass of the first gauge KK level should be at least $O$(3 TeV)~\cite{Agashe:2003zs,Contino:2006nn,Contino:2008xg}.  Incorporating constraints from flavor and CP can push the mass even higher~\cite{Gedalia:2010rj}.  These observations tend to generalize beyond this specific model, especially regarding electroweak corrections.  The implication is that if the Higgs is composite, then the physics associated with that compositeness may be hiding at the multi-TeV scale.  For this particular model, a 2(3) TeV $Z'$ can be produced in $q\bar q$ annihilation with $\sigma\times BR(Zh\ {\rm or}\ W^+W^-)$ of roughly 15($1.5$)~fb at a 14~TeV LHC~\cite{Agashe:2007ki}.  (This is, in fact, the dominant production mechanism.)

While we have so far discussed two extreme cases, namely simple gauge $U(1)'$ models and models where the Higgs and $Z'$ are both composite particles, it is also possible to interpolate between the behavior of these two.  A well-known example is the Little Higgs scenario~\cite{ArkaniHamed:2001nc} (reviewed in~\cite{Schmaltz:2005ky}).  Here, the Higgs is a composite pseudo-Goldstone boson arising from the breakdown of a partially-gauged global symmetry, and inherits special protection from one-loop mass corrections.  The gauge sector contains two copies of $SU(2)\times U(1)$ as a subgroup, of which our own electroweak group is a mixture.  The orthogonal gauge group contains two $Z'$ bosons.  Unlike the Higgs, they are now elementary, but they becomes massive due to the dynamics of the strongly-coupled sector that generates the Higgs.  The relative couplings of these $Z'$ bosons with the Higgs and with the SM fermions depends on the mixing angles between the two copies of the gauge group, and can be continuously dialed.\footnote{While the $Z'$ bosons here represent simple broken gauge symmetries, the branching fractions into fermions and $Zh$/\WpWm\ can escape the constraint of equation (\ref{BRratio}) since the Yukawa couplings in these models contain spurion fields to balance the charges.}  Similar to the warped 5D models, electroweak constraints favor TeV-scale masses or higher for the $Z'$ bosons of the Little Higgs (see, e.g.,~\cite{Csaki:2002qg,Csaki:2003si}).  To get a sense for what cross sections are possible, we cite the numbers for the ``Littlest Higgs''~\cite{ArkaniHamed:2002qy} (with cot$\theta = 0.5$) studied in~\cite{Atl:LH}:   10($1.1$)~fb for $m_{Z'} = 2(3)$~TeV at a 14~TeV LHC.

In summary, we emphasize that many well-motivated models contain one or more $Z'$ with appreciable coupling to the Higgs doublet, and therefore appreciable branching fraction into electroweak bosons.  Such $Z'$ are constrained by electroweak precision tests (as well as flavor measurements) to live in the multi-TeV mass range.  Through the analog of $\gamma$-$\rho^0$ mixing, it is possible to produce these resonances at the LHC through $q\bar q$ annihilation, even if the $Z'$ is a composite or belongs (mostly) to a gauge group that is decoupled from the light SM fermions.  In the case that the $Z'$ is a simple elementary gauge boson, we usually expect the branching fractions into electroweak bosons to be smaller than the branching fractions into fermions.  However, even if this turns out to be the case, verifying that the rate into electroweak bosons is either absent or subleading would be very useful to test the gauge-like nature of the $Z'$.

\subsection{Possible final-state decay channels}

While we have been discussing only two options for the initial decay stage of the $Z'$ (roughly equal rates into $Zh$ and \WpWm), the secondary decays of the electroweak bosons lead to a proliferation of possible final states.  Matters are particularly complicated by the fact that we do not yet know the mass of the Higgs boson, nor even if its branching fractions for a given mass strictly follow those predicted by the SM.

We can start by categorizing the final states from the well-understood \WpWm\ case.  The available topologies (identical those well-known from $t\bar t$) are dileptonic, $l$+jets, and all-hadronic.  (Dileptonic and $l$+jets include cases with taus, though we do not explicitly investigate these here.)  The dileptonic channel (5\% $BR$) is in principle the cleanest, but suffers from low branching fraction and ambiguities in kinematic reconstruction.  It was analyzed in~\cite{Agashe:2007ki} and found to be less sensitive than $l$+jets.  Given this, and given that the current paper is focused on jet substructure applications, we do not re-analyze this channel.  The $l$+jets channel (30\% $BR$) was also studied in~\cite{Agashe:2007ki} (as well as in~\cite{Benchekroun:2001je}), and was shown to be the most powerful.  We investigate the search reach in this channel in subsection~\ref{sec:WWlvjj}.  Finally, the all-hadronic channel (45\% $BR$) has high rate but faces overwhelming dijet background.  Even applying two hadronic $W$-tags to the event (using the substructure techniques of the next section) is not enough to combat this to the point where all-hadronic achieves a reasonable $S/B$ ratio.\footnote{Since the $l$+jets and all-hadronic branching fractions are comparable, all-hadronic can be competitive only if the single-tagged $W$+jet $\to (l\nu)$+jet and double-tagged dijet background rates are also comparable.  Equivalently, the ratio between the untagged background cross sections ($W$+jet/dijet) should be of order the $W$ mistag rate or larger.  We estimate this ratio of cross sections to be only $0.03$\% in our kinematic range of interest, more than two orders of magnitude too small, given the mistag rates we estimate in subsection~\ref{rates}.}   Therefore, we do not investigate it.

The $Zh$ case offers significantly more options.  We can globally classify these using the $h$ decay mode, and more locally by also specifying the $Z$ decay mode.

The simplest option, and the one favored by a light SM-like Higgs (70\% $BR$ for $m_h = 120$ GeV), is $h\to b\bar b$.  This two-body decay was explored in the highly boosted case in~\cite{Atl:LH,Agashe:2007ki}, and in the intermediate-boost case in~\cite{Butterworth:2008iy}.  The $b\bar b$ system will be accompanied by the decaying $Z$, which can decay to $l^+l^-$ ($l=e,\mu$, 7\% $BR$), neutrinos (20\% $BR$), $\tau^+\tau^-$ (3\% $BR$), or jets (70\% $BR$).  The usual logic is to focus on the cleanest possible final-state:  $(l^+l^-)(b\bar b)$, usually with one or two $b$-tags.  We will analyze this mode anew in subsection~\ref{sec:Zhllbb}, applying jet substructure and replacing vertex-based $b$-tagging, which performs uncertainly at high-$p_T$, with $\mu$-based tagging or no tagging at all.\footnote{While we do not explicitly exploit this, our no-tagging analysis will also capture the signal from $h\to gg$.}  The invisible $Z$ should be similarly clean, but has not been studied in detail before at high-$p_T$.  (It is mentioned in~\cite{Agashe:2007ki}, and studied explicitly at intermediate-boost in~\cite{Butterworth:2008iy}.)  While complete kinematic reconstruction is impossible, the $Z'$ mass can be approximated event-by-event by using $2p_T(h)$.  This has a Jacobian peak at $m_{Z'}$, which is further reinforced by spin terms in the decay matrix element.  While the resonance peak here is inevitably broader than in the fully reconstructed leptonic case, and is therefore subject to higher background, this interplays nontrivially with the three times higher statistics.  We study this mode in subsection~\ref{sec:Zhvvbb}.  Taus we will not consider, as they are challenging and have very small rate.  (However, we return to them in the context of Higgs decays below.)  Finally, the case with a jetty Z faces the same kind of overwhelming background problem as all-hadronic \WpWm, and is at best extremely challenging.\footnote{We might ask whether it could be recovered by demanding two hadronic electroweak boson tags and two $b$-tags on the Higgs side.  If the $b$-tag operates perfectly (100\% tag, 0\% mistag), this may not be so terrible.  However, it seems unlikely that this mode can be made competitive with more realistic $b$-tag numbers.}  We do not pursue it here.

This rounds out the set of modes which we explore in this paper:  $(l\nu)(q\bar q')$, $(l^+l^-)(b\bar b)$, $(\nu\bar\nu)(b\bar b)$.  We will pursue two other important options for Higgs decays, $h\to\tau^+\tau^-$ and $h\to WW^{(*)}/ZZ^{(*)}$, in a set of companion papers~\cite{ZprimesTau,Zprimes2}.

\section{Jet Substructure for Boosted Electroweak Bosons}
\label{sec:substructure}

%%%%%%%%%%%%%%%%%%%%%%%%%%%%%%
% Section on jet substructure
%%%%%%%%%%%%%%%%%%%%%%%%%%%%%%

The electroweak bosons produced in the decay of a multi-TeV $Z'$ will be highly Lorentz-boosted.  Consequently, their own decay products tend to become highly collimated into single ``jets'' --- $W$-jets, $Z$-jets, and $h$-jets.  On the one hand, this collimation can complicate searches, as these boson-jets can easily be confused with QCD-induced jets.  E.g., the mode $Z'\to WW \to (l\nu)(q\bar{q}')$ will suffer from large, quasi-two-body $W$+jets backgrounds $Wq$ and $Wg$.  Even modes with leptons within a jet (such as from boosted $h\to\tau^+\tau^-\to lX$), will face backgrounds from ordinary jets containing bottom/charm production, lepton fakes, or $W/Z$-strahlung.  On the other hand, the internal kinematic configurations of boson-jets are still rather distinctive from QCD, and we can capitalize on this to improve discrimination.  Indeed, a variety of strategies have been developed to discriminate hadronic (or semi-hadronic) decays of boosted heavy particles from ordinary QCD jets utilizing jet substructure, i.e. the detailed distribution of particles and/or energy within a jet.  In this section, we detail in what sense substructure techniques are beneficial for discriminating boosted $W/Z/h$ decaying to two quarks, and how robust these techniques are to contamination and detector effects.  (We discuss new techniques which we develop for the even more distinctive case of boosted $h\to\tau^+\tau^-$ in~\cite{ZprimesTau}, and boosted $h\to WW^{(*)}/ZZ^{(*)}$ in~\cite{Zprimes2}.)  

The section is organized as follows.  We first describe the basic utility of substructure methods and describe our nominal choice of algorithm.  We then make a comparison with a more traditional jet-based analysis to further clarify the advantages of substructure.  Finally, we estimate the performance of boosted electroweak boson tagging, and discuss the robustness of our estimates against theory and detector modeling.

\subsection{Utility of substructure and the BDRS algorithm}\label{BDRS}

The simplest jet observable sensitive to substructure is the jet's mass.  While the decay products of the boosted electroweak boson may end up within a single jet, this jet will still have a mass close to the electroweak boson mass, and the $Z'$ resonance will show up as a localized bump in the joint distribution of event mass and jet mass.  For backgrounds with QCD jets, the event mass and the jet mass will both tend to be peaked toward zero.

Taking an even closer look at the energy distribution within a jet offers two immediate improvements.  First, ordinary QCD jets with mass near, say, $m_W$, will preferentially acquire their mass via relatively soft emissions near the edge of the jet area (depending in detail on the jet definition and $p_T$ scale).  This situation is in contrast to genuine electroweak bosons, which usually consist of two well-localized subjets with roughly democratic energy-sharing.   Substructure techniques~\cite{Seymour:1993mx,Butterworth:2002tt,Butterworth:2008iy,Ellis:2009su,Almeida:2008yp,Almeida:2010pa} can identify the more QCD-like kinematic configurations and remove them from analysis, thereby improving signal-to-background.  Second, by identifying the subjets corresponding to the two decay partons, we can restrict our boson reconstruction to include only the particles (tracks and calorimeter cells) near their cores, thereby ignoring wider-angle radiation in the jet.  This tends to be dominated by uncorrelated sources, namely ISR, underlying event, and pileup.  By refining the set of particles to those in the immediate vicinity of the subjets, the jet mass resolution can be optimized.\footnote{A number of promising techniques have also been developed to specifically clean a jet of uncorrelated radiation, independent of the identification of localized subjets~\cite{Ellis:2009su,Krohn:2009th,Seymour:1993mx}.}

To take advantage of both of these improvements, we focus here on the subjet-finding technique developed by Butterworth, Davison, Rubin, and Salam (BDRS)~\cite{Butterworth:2008iy} and further investigated by ATLAS~\cite{ATLASBDRS}.  Our modes $Zh \to (l^+l^-)(b\bar b)$ and $Zh \to (\nu\bar\nu)(b\bar b)$ were also studied in those papers (as well as the closely related $Wh \to (l\nu)(b\bar b)$), at lower boosts and in the context of a Standard Model Higgsstrahlung search.  The BDRS technique trivially extrapolates to boosted hadronic $W$s and $Z$s, as well as to arbitrarily high energies, limited only by the detector's spatial resolution (a topic which we address in appendix~\ref{sec:detector}).

The algorithm runs as follows.  An event is first iteratively clustered into large-radius ``fat-jets'' with the Cambridge/Aachen (C/A) jet algorithm~\cite{Dokshitzer:1997in,Wobisch:1998wt}.   After a fat-jet of interest is identified, it is iteratively {\it declustered} by working backwards through the clustering stages.  At a given stage, a protojet $j$ is split into progenitors $j_1$ and $j_2$.  The splitting is categorized by a fractional mass-drop measure max$(m_1,m_2)/m_{12}$ and a symmetry measure min$(p_{T1}^2,p_{T2}^2)\Delta R_{12}^2/m_{12}^2$.  The algorithm stops, identifying two subjets, when the former is less than a threshold $\mu = 0.67$, and the latter is greater than a threshold $y_{\rm cut} = 0.09$.  If these criteria are not met, the more massive of $j_1$ and $j_2$ is used as the $j$ for the next iteration, and the less massive is thrown away.  The algorithm continues searching down to arbitrarily small $\Delta R$s and arbitrarily small protojet masses, until either suitable substructure is found or there are no more particles left to operate on.

While we will be applying this algorithm in a new kinematic regime, we do not modify its original declustering parameters.  Partially this is to maintain contact with the earlier literature, but more practically we do not find dramatic gains in signal/background discrimination by changing parameters.  Indeed, the fairly scale-free nature of the procedure leads to similar behaviors at all $p_T$ scales.  We do note that after completing our studies with the original BDRS algorithm, we found that the mass-drop criterion appears to play almost no role.  For all of our analysis samples, it is almost always strictly weaker than the symmetry criterion.\footnote{Analyses run with the mass-drop criterion turned off yield kinematic distributions (both shape and normalization) for signal and background subjets that are nearly identical to analyses run with the mass-drop turned on.  However, running with the mass-drop but without the symmetry criterion leads to significant reshaping of background, and in particular enhances the number of background events in the electroweak boson mass windows by a factor of $2\sim3$.}  Of course, a more realistic study should more closely explore optimization of the procedure and its parameters. 

For our clustering radius, we use $R = 1.4$, corresponding to quasi-hemispheric fat-jets.  This may seem unnecessarily, even counterproductively, large for a kinematic region where the interesting structure becomes smaller than a normal jet.  However, we have found that this very large jet radius offers a number of advantages.  First, choosing a jet radius based on the known ratio $m_W/p_T$ or the expected ratio $m_h/p_T$ would introduce a hidden dimensionful scale into the analysis, manifesting itself as a turn-off feature beyond $m_W$ or $m_h$ in the background's jet-mass spectrum.  Choosing a large $R$ pushes this feature away from the physical boson mass, flattening out the sidebands.  Second, using a large $R$ broadens the range of $Z'$ masses over which our search could operate efficiently, down to $m_{Z'} = 300\sim400$ GeV.  Third, as discussed in subsection~\ref{rates}, using such a large fat-jet radius picks up on more global event information, which appears to be useful for discriminating electroweak boson jets from QCD jets.

After finding subjets, BDRS further refine them using a technique called filtering.  The set of particles constituting the two subjets found by the mass-drop/asymmetry procedure are reclustered with C/A, using a refined jet-radius parameter $R_{\rm filt} = {\rm min}(0.3,\Delta R_{\rm subjets}/2)$.  The hardest three of these refined subjets is kept and used to reconstruct the boson.  Filtering has been shown to be important to achieve optimal mass resolution at moderate boost, where the initial two subjets can be quite large and quite contaminated by the underlying event and pileup (see~\cite{Rubin:2010fc} for a detailed study).   As we transition to high boost, these subjets become smaller, and the effects of contamination quickly become much less dramatic.  For simplicity, we do not use filtering in the present study, nor do we introduce pileup into our simulation samples.  We separately address the utility of filtering for removing pileup (mainly relevant for the 1 TeV samples) in appendix~\ref{sec:pileup}.

Before proceeding, we also point out the possible utility of jet ``superstructure''~\cite{Gallicchio:2010sw} for discriminating boosted electroweak bosons from QCD jets.  The shower products of a boosted electroweak boson tend to be very well-collimated, at the $\Delta R$ scale of the two daughter quarks, as there is no color connection with the rest of the event.  QCD jets, on the other hand, can shower at all angles, often with bias towards one of the beams.  This can lead to differences in the large-angle energy patterns.  While simple measures of this difference lead to only modest discrimination~\cite{Gallicchio:2010sw,Hatta:2009nd}, it is possible that a more sophisticated analysis could yield a more powerful discriminator~\cite{Zhenyu}.  We will see in subsection~\ref{rates} that our choice of quasi-hemispheric jet clustering already incorporates some aspect of this more global discrimination.

%We use a clustering radius $R = 1.4$, corresponding to quasi-hemispheric fat-jets.\footnote{The detailed choice of $R$ is not crucial, especially as $\Delta R\ll 1.4$ for electroweak boson decays in the kinematic range explored in this paper.  However, we have explicitly chosen a very large radius for four reasons.  First, $R\times p_T$ constitutes a dimensionful scale which will manifest itself as a broad turn-off feature in the background's jet-mass spectrum.  Choosing $R$ as large as possible pushes this feature far away from the physical boson mass.   Second, using a large $R$ broadens the range of $Z'$ masses over which our search could operate efficiently, down to $m_{Z'} = 300\sim400$ GeV.  Third, large $R$ demonstrates the extreme stability of the algorithm, even in the presence of ISR and other uncorrelated hadronic activity.  Fourth, as discussed in subsection~\ref{rates}, using such a large fat-jet radius picks up on more global event information, which appears to be useful for discriminating electroweak boson jets from QCD jets.} 

\subsection{Comparison with traditional jet analysis}\label{monodijets}

We attempt to further clarify the utility of the BDRS procedure in heavy $Z'$ searches by performing a simple particle-level comparison with a more traditional jet-based analysis.  Analyses of this latter type have previously been investigated in~\cite{Benchekroun:2001je,Atl:LH,Agashe:2007ki}.  We use for this comparison events from our $Z'\to WW \to (l\nu)(q\bar{q}')$ signal and combined $W(q/g)$ background.\footnote{Details of the simulations and reconstruction can be found in section~\ref{sec:WW}, though there are some important differences in the present analysis.  No detector model is applied, and particle energies are not smeared.  (Including these effects would not significantly alter our conclusions.)  Also, unlike the full $WW$ analysis in section~\ref{sec:WW}, a jet veto is not applied here, as the sample of jets in the two analyses are very different, and we do not consider top backgrounds.}  Our BDRS-style analysis operates on the hardest $R=1.4$ fat-jet found in each event.  The traditional jet analysis clusters the event into $R=0.4$ anti-$k_T$ jets.  To make the two analyses as similar as possible, in the traditional analysis we first find the hardest jet and then seek out the next-hardest jet within $R=1.4$.  If the pairing is not too asymmetric, in the sense of the BDRS $y_{\rm cut}$, these two jets are kept as the ``subjets'' constituting the hadronic $W$ candidate.  (As in the substructure analysis, the mass-drop criterion is almost always satisfied.)  Failing this, it is assumed that the $W$ decay products have merged into one jet, and only the initial hardest jet is kept.  In both analyses, we only consider relatively central hadronic $W$ candidates, with $|\eta| < 1.5$.

\begin{figure}[tp]
\begin{center}
\epsfxsize=0.44\textwidth\epsfbox{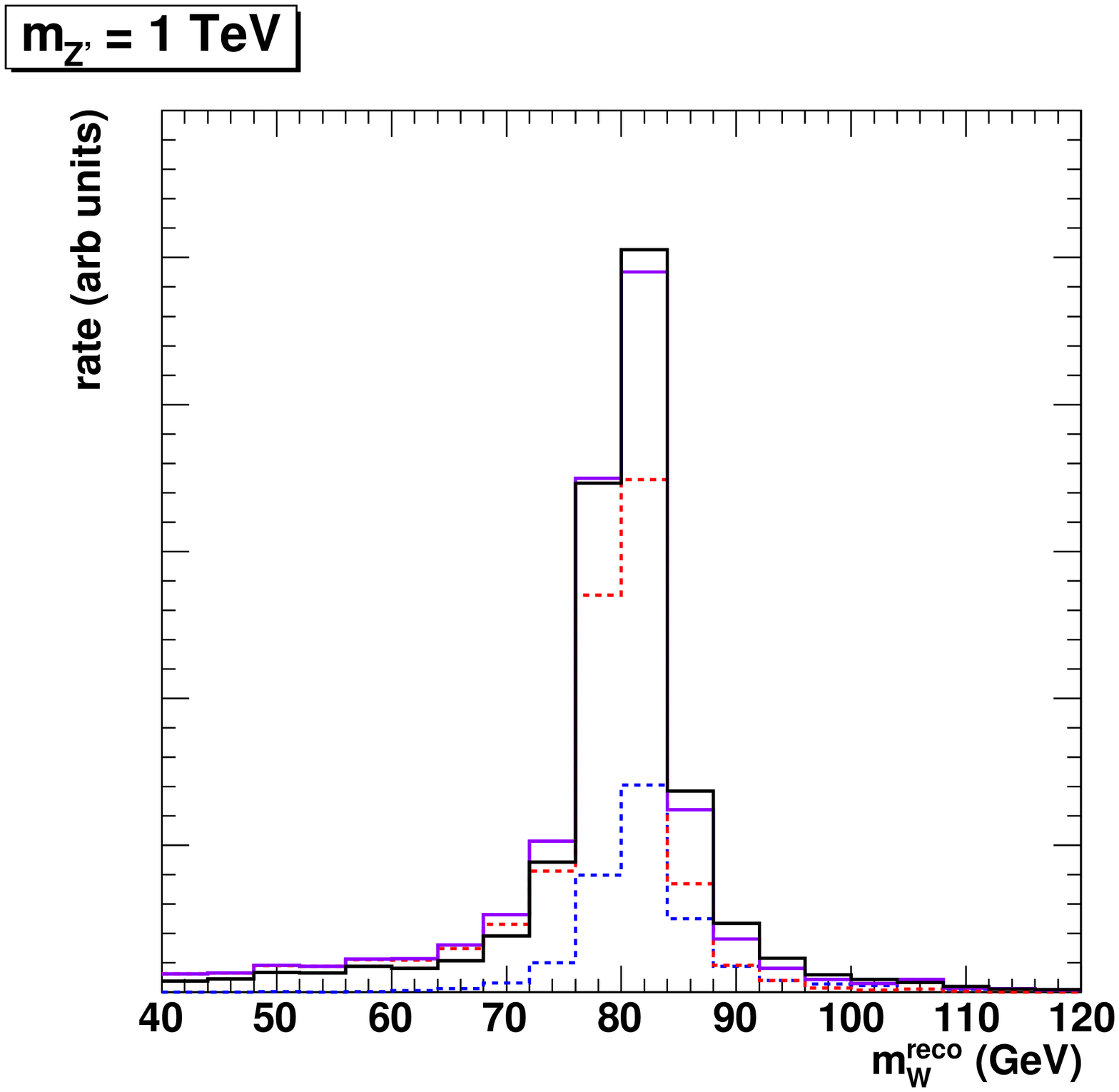}
\epsfxsize=0.44\textwidth\epsfbox{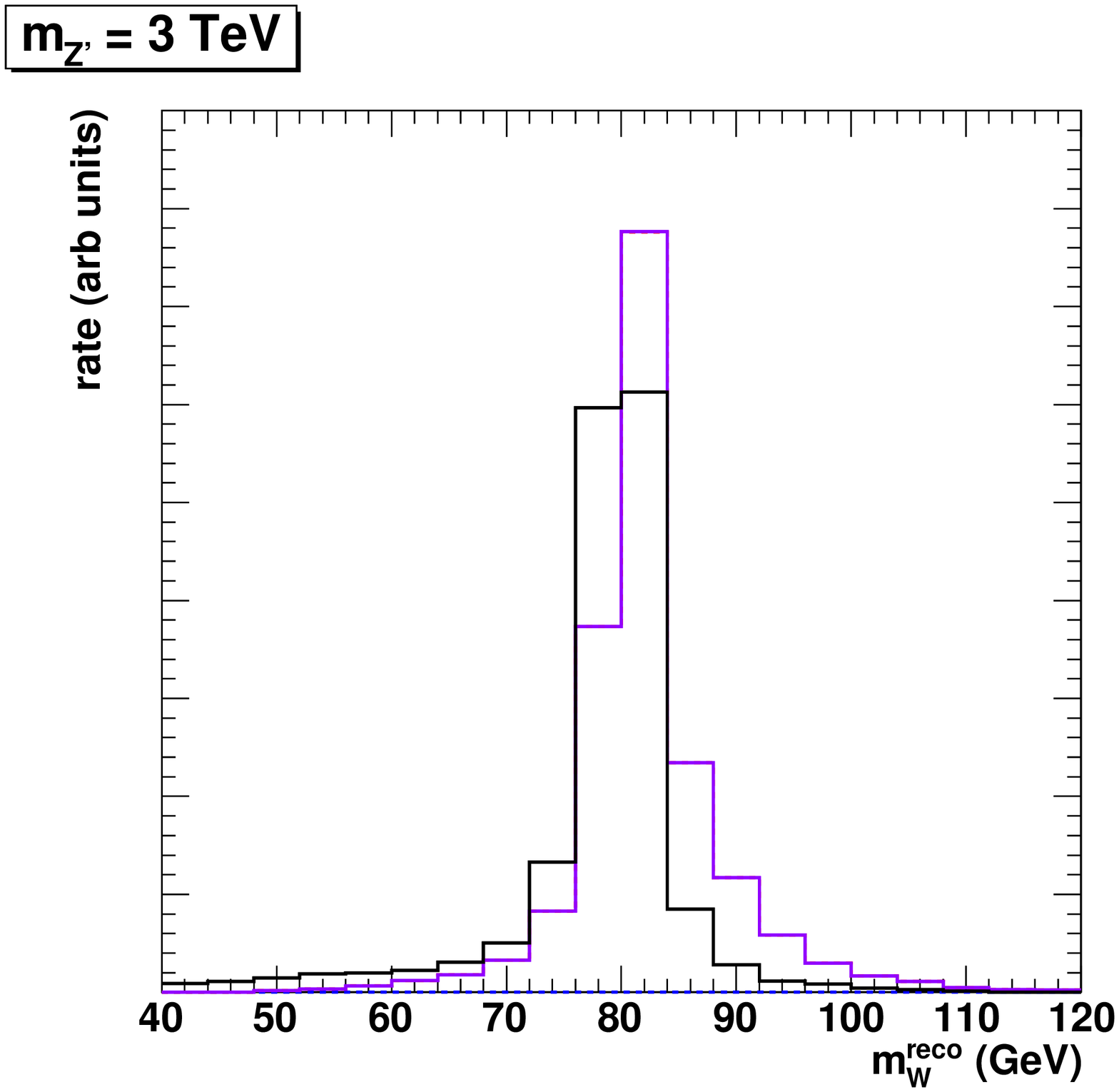}
\caption{\it Distributions of the (particle-level) reconstructed hadronic $W$ mass for 1 and 3 TeV $Z'\to WW \to (l\nu)(q\bar{q}')$.  Displayed are the nominal BDRS procedure (solid black), traditional dijets (dashed blue), traditional monojets (dashed red), and combined dijet+monojet (solid purple).}
\label{fig:monodijetSignal}
\end{center}
\end{figure}

The hadronic $W$ mass reconstruction is shown in Fig.~\ref{fig:monodijetSignal} for 1 and 3 TeV $Z'$.  It is clear that the impact of switching to a traditional analysis is modest.  Three features are worth noting.  First, at 1 TeV we see a healthy admixture of traditional dijet and monojet reconstructions, whereas at 3 TeV monojet (unsurprisingly) dominates.  Second, the traditional analysis is somewhat more efficient (by about $1.5\%$ at 1 TeV and $7\%$ at 3 TeV), due to the lack of internal phase space cuts in the monojet mode.  Third, the 3 TeV monojet sample shows some broadening due to underlying event contamination, since the separation between $W$-subjets is becoming small compared to the jet radius.

\begin{figure}[tp]
\begin{center}
\epsfxsize=0.44\textwidth\epsfbox{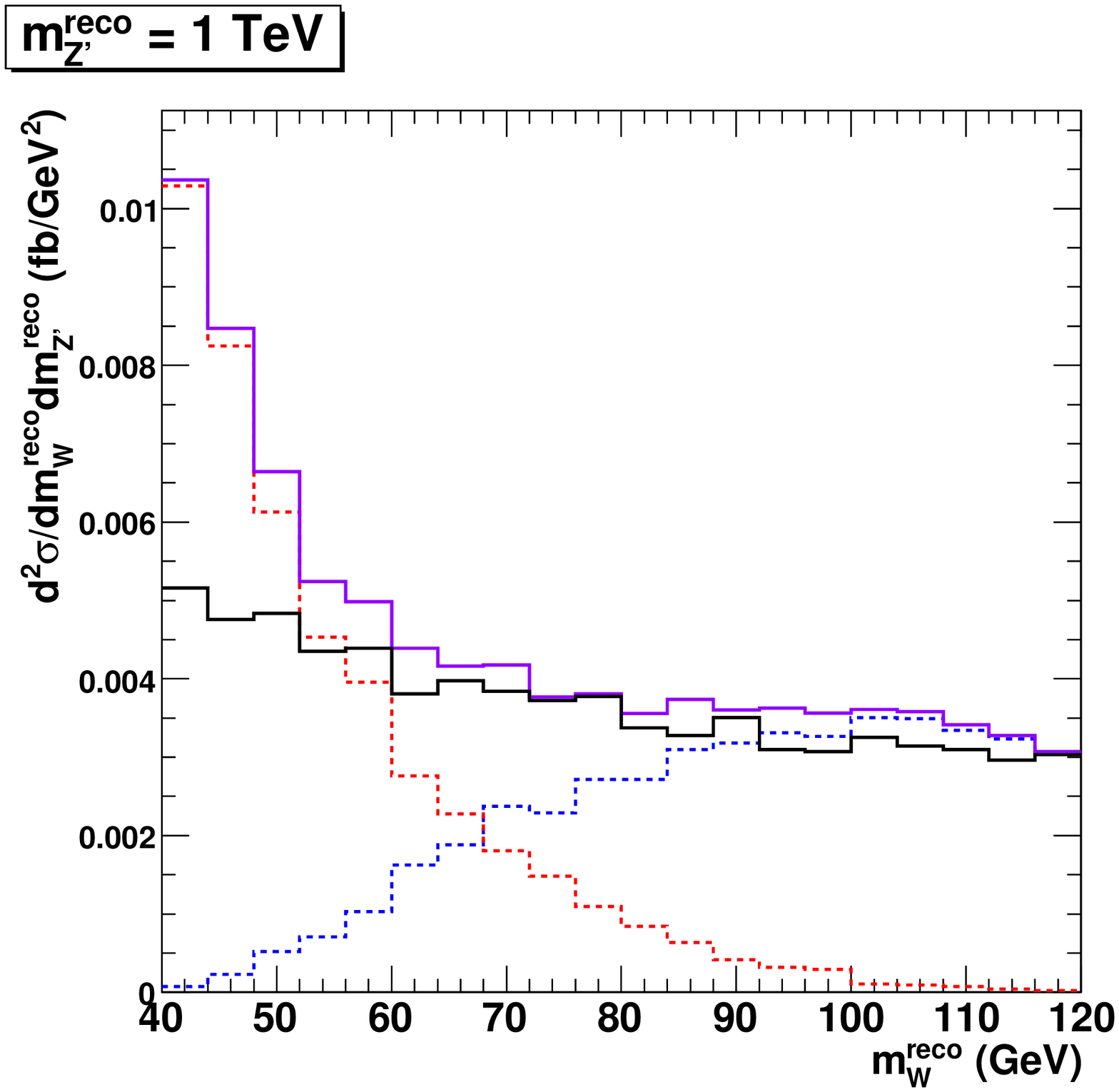}
\epsfxsize=0.44\textwidth\epsfbox{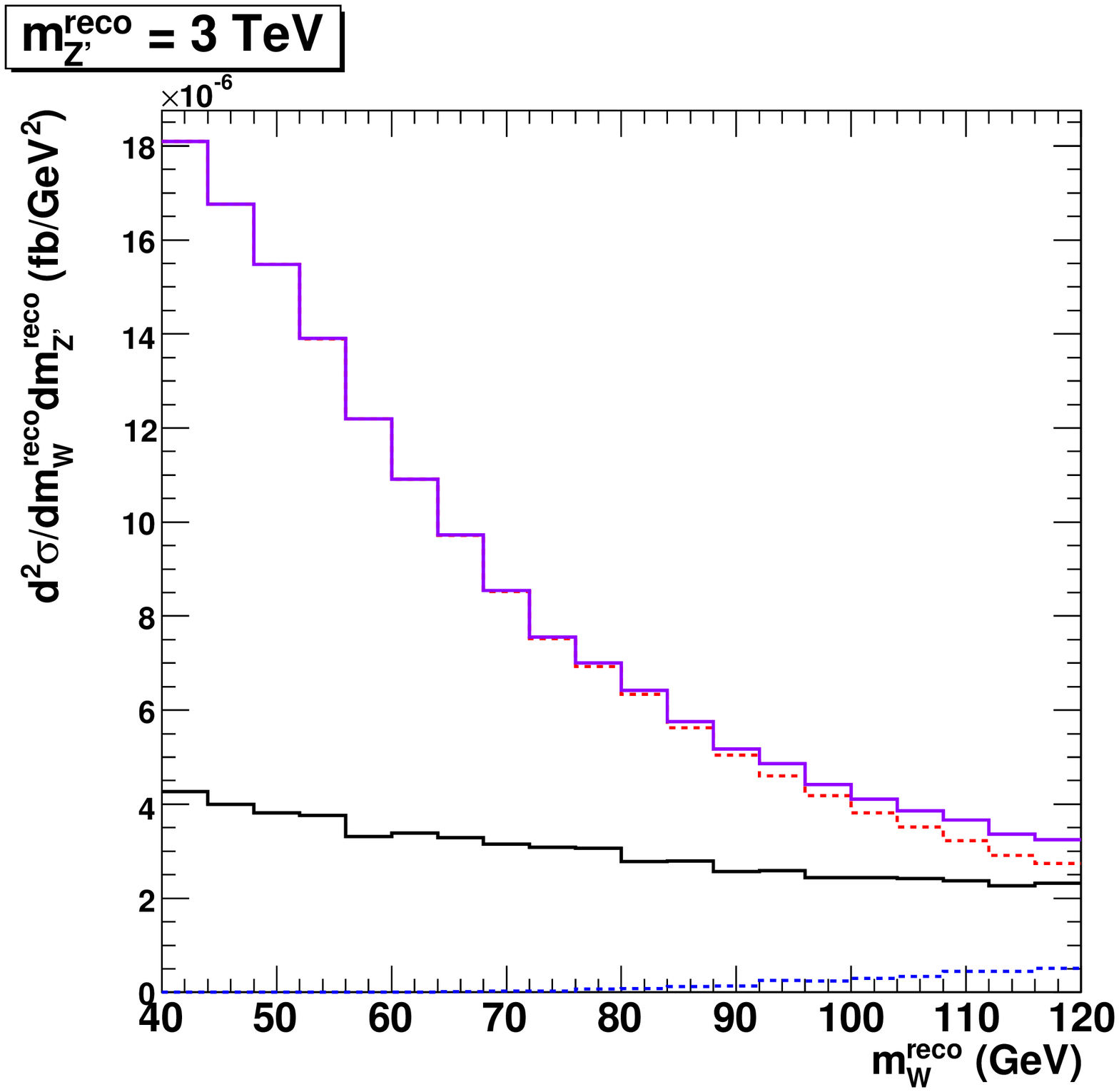}
\caption{\it Slices of the showered leading-order double-differential cross section for $W$+jets $\to(l\nu)$+jets vs (particle-level) reconstructed $W$ mass, fixing the reconstructed $Z'$ mass at 1 and 3 TeV.  Displayed are the nominal BDRS procedure (solid black), traditional dijets (dashed blue), traditional monojets (dashed red), and combined dijet+monojet (solid purple).}
\label{fig:monodijetBackground}
\end{center}
\end{figure}

The backgrounds are shown in Fig.~\ref{fig:monodijetBackground}, expressed as slices of the double-differential cross section in the $(m_W^{\rm reco},m_{Z'}^{\rm reco})$ plane at $m_{Z'}^{\rm reco} = 1$ and 3 TeV.  Again, 1 TeV events reconstructed near the $W$ mass are a mixture of traditional dijet and monojet, and each has a very distinct shape.  In the $W$ mass region, the sum is nonetheless quite comparable to the BDRS result.  At 3 TeV, the traditional analysis is again dominated by monojet.  However, the effects of the internal kinematic cuts of the BDRS procedure are now evident, as the background processed through BDRS is roughly two times smaller, and significantly flatter.

While the advantage of BDRS over more traditional jet analysis is very clear in the 3 TeV background, the 1 TeV background illustrates a more subtle potential advantage.  In the traditional analysis, we inevitably face the issue of combining dijet and monojet analyses.  There is a broad crossover region between dijet-dominance and monojet-dominance as we scan through potential $Z'$ masses.\footnote{The detailed behavior of this crossover depends on the jet definition and cuts.  For example, when we focus on the region $m_{Z'}^{\rm reco} = 1$ TeV, the reconstructed $W$s will have $p_T \sim 500$ GeV.  In our dijet analysis, the minimum $\Delta R$ is $0.4$, and the minimum $y$ is $y_{\rm cut} = 0.09$.  The minimum dijet mass can be roughly approximated as $\sqrt{0.09}(0.4)(500\;{\rm GeV}) = 60$ GeV, which corresponds reasonably well with the left panel of Fig.~\ref{fig:monodijetBackground}.  Note that changing (or removing) the symmetry cut will simply move the location of the crossover.  Taking a larger $R$ for the jet is another option.  However this subjects the reconstructed $W$s to more contamination, and is less discriminating against highly asymmetric QCD configurations.}  In contrast, the BDRS procedure is largely free of any distance or energy scales beyond the fat-jet clustering size.  It operates smoothly and (mostly) without dimensionful thresholds.  This feature may be advantageous, both in simplifying the analysis and in avoiding possible systematic effects on the background shape in the $(m_W^{\rm reco},m_{Z'}^{\rm reco})$ plane.  The relative flatness over $m_W^{\rm reco}$ may also be seen as an advantage.  The background processed by BDRS is only steeply falling along the $m_{Z'}^{\rm reco}$ direction, possibly allowing a signal bump to stand out more clearly in the mass plane than in the more traditional analysis.

\subsection{Performance and robustness}\label{rates}

\begin{figure}[tp]
\begin{center}
\epsfxsize=0.44\textwidth\epsfbox{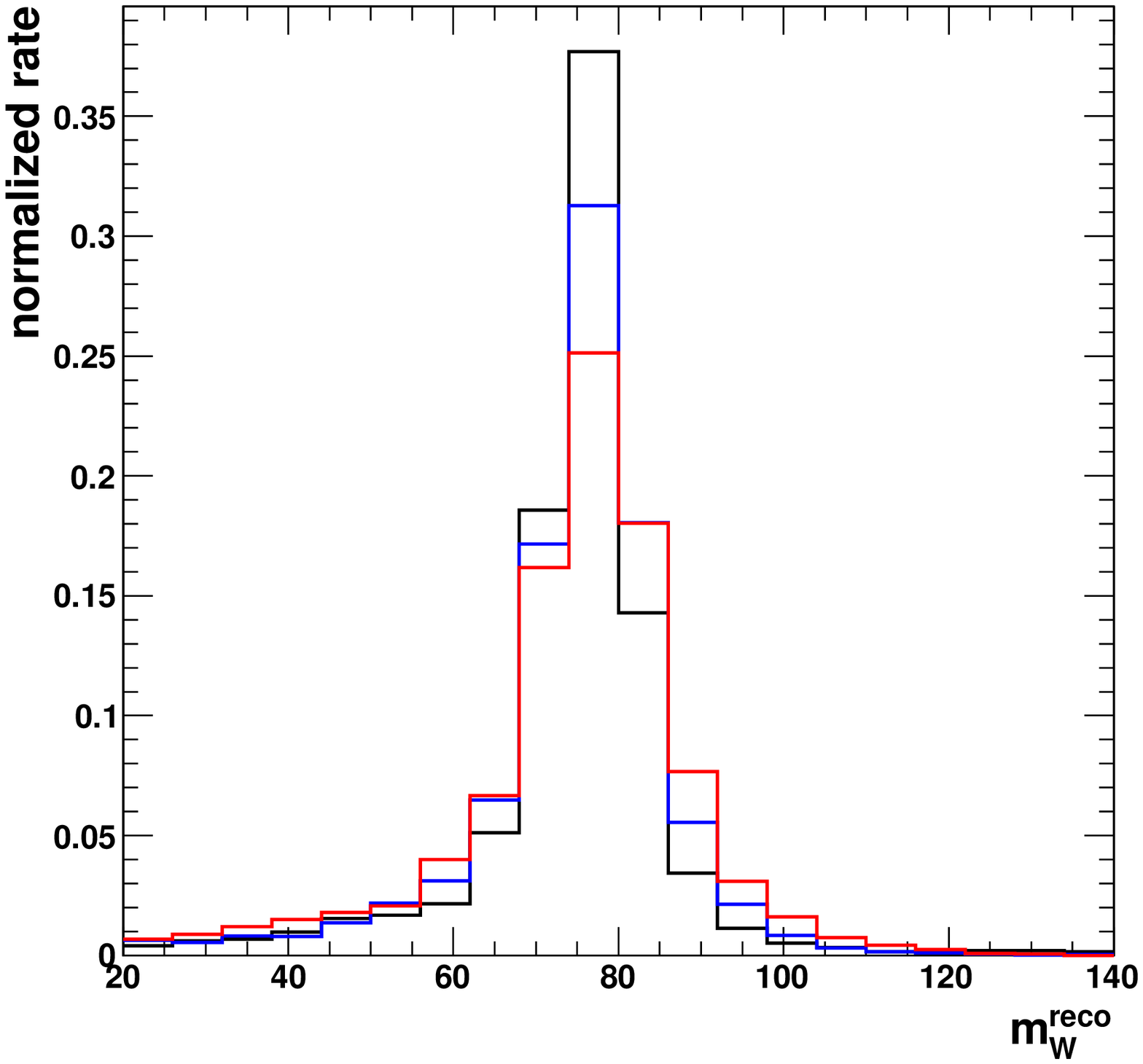}
\epsfxsize=0.44\textwidth\epsfbox{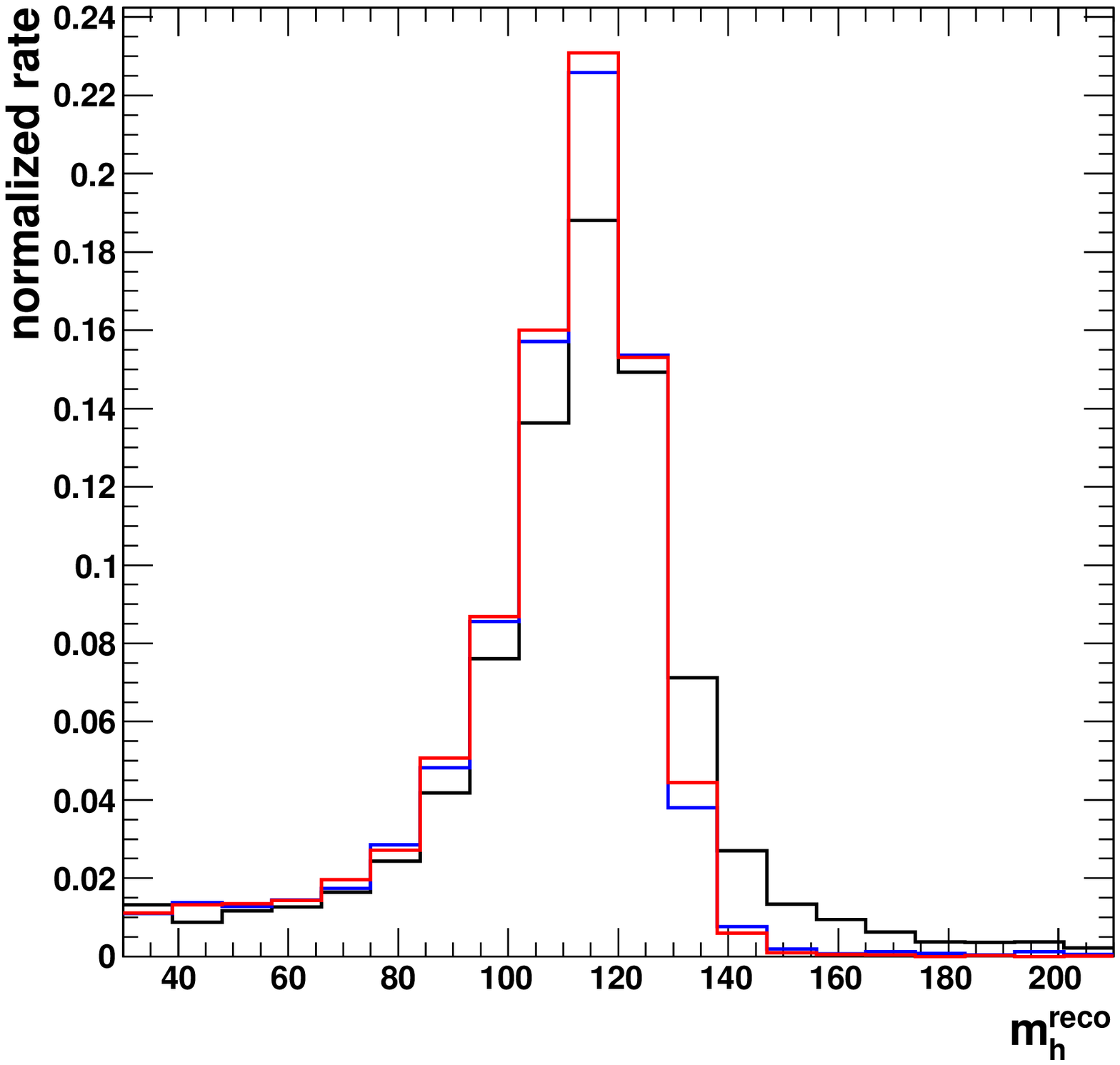}
\caption{\it Reconstructed $W$-jet (left) and $h$-jet (right) invariant mass distributions for $Z'$ masses of 1 TeV (black), 2 TeV (blue), and 3 TeV (red).  The rescaled ECAL model of appendix~\ref{sec:detector} has been applied.}
\label{fig:mass_peaks}
\end{center}
\end{figure}

\begin{table}[tp]
 \centering
\begin{tabular}{ c|c c c }
             &  $\;\; p_T \simeq 500$ GeV $\;\;$ & $\;\; p_T \simeq 1000$ GeV $\;\;$ & $\;\; p_T \simeq 1500$ GeV $\;\;$  \\ \hline
$W$\         &                           76\%    &                                77 &                               72  \\
$h$\         &                           59\ \ \ &                                61 &                               62  \\
\end{tabular}
\caption{\it Tag rates (in percent) for boosted electroweak bosons.  (Absolute statistical errors are of order 2\%.)}
\label{tab:eff_s}
\end{table}

We now categorize the performance of the BDRS procedure for discriminating boosted $W/Z/h$.  To make our estimates somewhat more realistic, we apply a simple detector model, which is described in detail in appendix~\ref{sec:detector}.  This model hybridizes ECAL spatial resolution with ECAL+HCAL energy measurements, and introduces rudimentary energy smearing.   The simple trick of using the ECAL to trace the energy flow significantly decreases the impact of detector granularity at very high $p_T$.  Event reconstruction follows section~\ref{sec:WW}.

For the signal, we use $W$-jets and (non-$b$-tagged) $h$-jets reconstructed from the $Z' \to WW \to (l\nu)(q\bar q')$ and $Z' \to Zh \to (l^+l^-)(b\bar b)$ samples, respectively.  (We do not present explicit numbers for $Z$-jets, but these will be very similar to $W$-jets.)  The boosted bosons produced in the $Z'$ decays are strongly peaked toward $p_T = m_{Z'}/2$, and are usually quite central.  The tag rate is defined as the fraction of reconstructed events where the summed boson subjets fall within a prespecified mass window chosen by eye:  $m_{\rm subjets} = [65,95]$ GeV for $W$s, and $m_{\rm subjets} = [100,140]$ GeV for Higgses.\footnote{Sometimes a boosted boson will fail reconstruction because the BDRS procedure deconstructs its fat-jet down to a single calorimeter cell.  We have checked that this is very rare for signal, at the few percent level at most.  It can be more important for QCD fat-jets, which tend to have less substructure.  The loss rate in that case is still typically less than 10\%.  Effectively, these represent an additional contribution to the zero-mass bin which we have not accounted for.}  The reconstructed $W$ and Higgs mass distributions are shown in fig.~\ref{fig:mass_peaks}, and tag rates are displayed in table~\ref{tab:eff_s}.  Qualitatively, we can see that the rates are $O(1)$, are weak functions of $p_T$, and that the Higgs rates tend to be smaller.  This last feature comes from the fact that the higgs is scalar, versus a longitudinal spin-1 boson, and that the mass peak is slightly degraded by neutrinos from semileptonic bottom and charm decays.  It is also worth noting that the the tag rates for transverse $W$ would be smaller than those for longitudinal, though we have not investigated this quantitatively.\footnote{Determining this would be important, for example, for $W$ bosons produced in decays of spin-0 or spin-2 resonances.}

\begin{table}[tp]
 \centering
\begin{tabular}{ c|c c c }
                &  $\;\; p_T \simeq 500$ GeV $\;\;$ & $\;\; p_T \simeq 1000$ GeV $\;\;$ & $\;\; p_T \simeq 1500$ GeV $\;\;$  \\ \hline
quark $\to W$\  &                       6.5\%        &                               6.5 &                              5.9  \\
quark $\to h$\  &                       6.8\ \ \     &                               5.6 &                              5.8  \\ \hline
gluon $\to W$\  &                      10.4\ \ \ \ \ &                               8.3 &                              7.4  \\
gluon $\to h$\  &                      10.5\ \ \ \ \ &                               8.8 &                              7.4  \\
\end{tabular}
\caption{\it Mistag rates (in percent) for quarks and gluons faking boosted electroweak bosons.  (Absolute statistical errors are of order $0.3$\%.)}
\label{tab:eff_b}
\end{table}

The mistag rates are displayed in table~\ref{tab:eff_b}, for $Z(q/g)$ backgrounds generated in $\hat{p}_T$ bins of $[400,600]$ GeV, $[800,1200]$ GeV, and $[1200,1800]$ GeV.  ($W(q/g)$ simulations yield identical results.)  These rates are also fairly weak functions of $p_T$, and are similar for our $W$ and Higgs mass windows.  Quarks fake heavy bosons less often than gluons, as their probability for radiating is smaller.  The final numbers are about 6-7\% for quarks, and  7-11\% for gluons.

Interestingly, the mistag rates show a decreasing trend as we increase $p_T$, particularly for gluons.  This may be indicating some of the global color effects mentioned in subsection~\ref{BDRS}.  In particular, our very large fat-jet size will pick up FSR emissions at wide angles, which are present for QCD jets but not for electroweak boson jets.  This hardens the background jet-mass spectrum, sending some fraction of background events beyond the electroweak boson mass windows.  To get a better sense for what this implies, we re-ran the $W$-tag analysis on the $m_{Z'} = 3$ TeV $WW$ signal and $p_T = [1200,1800]$ backgrounds with $R=0.3$.  This jet size is just large enough to contain the subjets from almost all of the signal events within our $W$ mass window, given the value of $y_{\rm cut}$.  We find that the signal efficiency increases slightly, from 72\% to 74\%.  This represents some recaptured events, which were lost in the quasi-hemispheric analysis due to high-mass combinations where one subjet is the entire boosted $W$ and the other is an ISR jet.  (The original fraction of such ``bad'' combinations is at the 5\% level for all $Z'$ samples.)  However, the quark and gluon mistag rates increase by a factor of roughly $1.7$, to $9.4$\% and $13.0$\%, respectively.  Optimization of the fat-jet and subjet-finding procedures to best capitalize on this physics is an interesting topic which we defer to future work.  Still, it is already clear that using smaller jets at higher $p_T$ may, counterintuitively, be {\it detrimental}.

We have also investigated the theoretical robustness of these mistag rate estimates by comparing our results against samples generated with {\tt HERWIG v6.510}~\cite{Corcella:2000bw} and {\tt MadGraph/ MadEvent+PYTHIA \tt v4.4.44}~\cite{Alwall:2007st} with jet-parton matching.  The {\tt HERWIG} samples are of identical composition to the default {\tt PYTHIA} samples.  They display no statistically significant difference in any of the mistag rates.  The {\tt MadGraph} sample consists of $Z$+jets events MLM-matched~\cite{Hoche:2006ph} up to three jets using $p_T$ and $\Delta R$ thresholds of 30 GeV and $0.2$, respectively.  We have generated the $p_T = [400,600]$ GeV bin, and find rather good agreement with the {\tt PYTHIA} and {\tt HERWIG} results.\footnote{When working at higher-order it becomes somewhat ambiguous whether a jet originates from a quark or a gluon.  As an operational definition, we search out hard progenitor partons within $\Delta R=1.4$ of the main fat-jet axis, tallying individual (signed) quark flavors.  Jets with leftover quark-number are deemed ``quark jets,'' and are otherwise deemed ``gluon jets.''  This procedure yields relative ``$Zq$'' and ``$Zg$'' cross sections quite comparable to what is obtained with simple $2\to2$ matrix elements.}  The mistag rates from {\tt MadGraph} do appear to be 10-20\% lower, relatively speaking, but this would have to be confirmed with higher-statistics running.

Detector modeling (see appendix~\ref{sec:detector}) is largely unimportant for the mistag rates given fixed jet-mass windows, as the QCD spectrum processed through BDRS is not sharply featured.  We have explicitly checked that eliminating the calorimeter geometry model and/or the energy smearing has no significant effect.  The jet-mass distributions for signal are of course much more sensitive.  Practically, we can parametrize our ignorance with a single $p_T$-dependent factor:  the amount by which we have mis-estimated the jet-mass resolution.  If a more realistic estimate tells us that we must expand our mass windows by a factor $x$ to accept the same signal, then our mistag rates will increase by approximately the same factor (since the background spectra are fairly flat).  Signal significance will scale like $1/\sqrt{x}$, and signal-to-background like $1/x$.  Assuming that $x$ is an $O(1)$ number, then our estimates for $Z'$ reach should also be good to $O(1)$.

\section{$Z'$ Searches}
\label{sec:WW}

%%%%%%%%%%%%%%%%%%%%%%%%%%%%%%%%%%%%%%%%%%%%%%%
% Section on Z'->(lv)(jj), (ll)(bb), (vv)(bb)
%%%%%%%%%%%%%%%%%%%%%%%%%%%%%%%%%%%%%%%%%%%%%%%

In this section, we estimate the discovery reach for $Z'$ resonances decaying to \WpWm\ and $Zh$, applying the jet substructure techniques in section~\ref{sec:substructure}.  As discussed in section~\ref{sec:theory}, $l$+jets is the most promising final state for \WpWm, and we investigate this mode.  This has previously been studied in~\cite{Benchekroun:2001je,Agashe:2007ki}.  For $Zh$, we focus exclusively on the case of a light Higgs (120 GeV) decaying to \bbbar.  The channel with associated leptonic $Z$ decay has traditionally been considered the most promising~\cite{Atl:LH,Agashe:2007ki} (see also~\cite{Hackstein:2010wk}), and we also investigate this mode.  The channel with invisible $Z$ has been mentioned in~\cite{Agashe:2007ki}, but not investigated in detail.  This is the final mode which we study here.  Perhaps not without coincidence, these three topologies --- $W^+W^-\to(l\nu)(q\bar q')$, $Zh\to(l^+l^-)(b\bar b)$, and $Zh\to(\nu\bar\nu)(b\bar b)$ --- are nearly the same as those studied in~\cite{Butterworth:2008iy}, though with $Wh$ replaced by $WW$.\footnote{We note that our results for $WW$ can be translated with minor modifications to the case of $W'\to Wh\to(l\nu)(b\bar b)$ without $b$-tags.  Our non-tagged $Zh$ results can also be translated for searches for $ZZ\to(l^+l^-)(q\bar q)$ and/or $ZW\to(l^+l^-)(q\bar q')$, though some care should be taken in accounting for spin effects.}

The $Z'$ samples for this study are generated with 2$\to$4 matrix elements in {\tt MadGraph/ MadEvent v4.4.32}~\cite{Alwall:2007st} and showered/hadronized in {\tt PYTHIA} (plugin version {\tt 2.1.3}).  The $Z'$ width is taken to be 3\%, which is smaller than instrumental width and the signal windows that we use below.  (Our estimates should work reasonably well for resonances up to about 15\% natural width.)  Our background samples are generated with {\tt PYTHIA v6.4.11}~\cite{pythiamanual} with default settings.\footnote{{\tt PYTHIA} accounts for spin effects in the electroweak boson decay distributions for these backgrounds.  However, they would not be accounted for in $Z'$ decays.}  All processes are evaluated at leading-order.\footnote{The NLO QCD $K$-factor for the $Z'$ signal is $1.3$~\cite{Fuks:2007gk}.  The QCD $K$-factor for high-$p_T$ $W/Z$+jets, which is our dominant background in all cases, is $1.6$~\cite{Rubin:2010xp}.  The net effect of these corrections on $S/\sqrt{B}$ is therefore small.  Electroweak corrections for high-$p_T$ processes can also be significant~\cite{Kuhn:2007cv}.  These tend to reduce the $W/Z$+jets rate.}  As discussed in appendix~\ref{sec:pileup}, we do not incorporate pileup into the analysis, but have checked that its effects would be quite manageable.  The final-state particles are processed through a primitive ECAL/HCAL hybrid calorimeter model, described in detail in appendix~\ref{sec:detector}, to capture the main spatial resolution effects which we face in reality.  The resulting calorimeter cells are then clustered/declustered using {\tt FastJet 2.4.1}~\cite{Cacciari:2005hq}.  Energy smearing is applied to the final reconstructed objects (leptons and subjets).  Subjet energy smearing is described in appendix~\ref{sec:detector}.  Electrons are smeared by 2\%, and muons by $(5\%)\sqrt{E/{\rm TeV}}$.

We estimate discovery reach for 1, 2, and 3 TeV $Z'$ by using simple counting.  For each sample, we construct a box in the $(m_{W/h}^{\rm reco},m_{Z'}^{\rm reco})$ plane, centered on the signal.  The box is not optimized, other than coarsely by eye.  We claim that discovery is possible if two criteria are met:  $N_S/\sqrt{N_B} > 5$ and $N_S > 10$.

Several of our reconstruction criteria are common between all of the analyses.  We accept leptons ($e$ and $\mu$) with $p_T > 30$ GeV and $|\eta| < 2.5$.  The leptons are required to be isolated from surrounding hadronic activity (neglecting photons and other leptons) within a cone of $R=0.4$:  $p_T(l)/(p_T(l)+p_T({\rm hadrons})) > 0.9$.  Jets are constructed, as described in section~\ref{sec:substructure}, using the Cambridge/Aachen algorithm with $R=1.4$.  We consider jets with $p_T > 20$ GeV and $|\eta| < 2.5$.  The leading jet, which we take to be the hadronic electroweak boson candidate, must have $p_T > 200$ GeV and $|\eta| < 1.5$.  (The last requirement forces the candidate into the better-segmented barrel region of the detector.)  This jet is declustered, and the two subjets used to reconstruct the boson.

%--------------------------------------------------------
\subsection{\WpWm\ in $l$+jets mode: $(l\nu)(q\bar q')$}
\label{sec:WWlvjj}
%--------------------------------------------------------

\begin{figure}[tp]
\begin{center}
\epsfig{figure=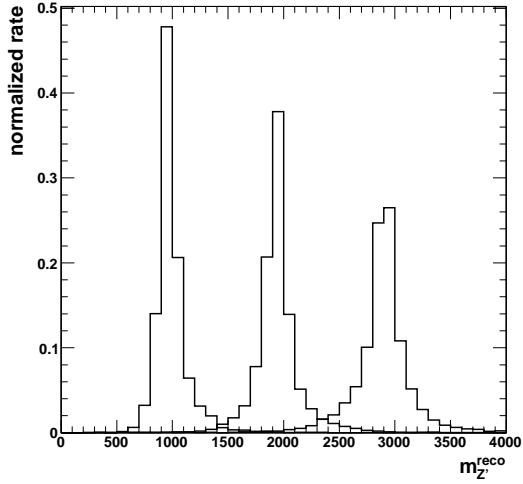,scale=0.37}
\caption{\it Reconstructed $Z'$ invariant masses for 1, 2, and 3 TeV in $WW \to (l\nu)(q\bar q')$.}
\label{fig:WWresonance}
\end{center}
\end{figure}

Our basic selection cuts are as follows.  We demand exactly one isolated lepton, which should be located in the semicylinder opposite the hadronic $W$ candidate jet.  We model missing energy by taking the transverse vector that balances the (energy-smeared) lepton and two subjets.  We then approximately reconstruct the complete leptonic $W$ boson by exploiting the quasi-collinearity of the lepton and neutrino:  $\eta_\nu \equiv \eta_l$.  After this step, we can construct the complete event mass, $m_{Z'}^{\rm reco}$.  To reduce $W$+jets background, which tends to be produced at shallower angles than the signal for a given $\hat s$, we demand that the hadronic $W$ satisfy $p_T > m_{Z'}^{\rm reco}/3$.  Finally, in order to control top backgrounds, we apply a jet veto:  the subleading jet in the event (if found) should satisfy $p_T < p_T(l)/2$.\footnote{This is a very simplistic cut, with many caveats.  In particular, since the jets we use are very large, pileup can significantly increase the subleading jet's momentum.  However, even a slightly more sophisticated analysis could likely work around this, for example reclustering with a smaller jet size or trimming the fat-jet~\cite{Krohn:2009th}.  We suspect that a more dedicated top ``anti-tag,'' for example attempting to fully reconstruct and veto the leptonic and/or hadronic top, could perform even better.  We also note that a jet veto of this type rejects $W$+jets backgrounds generated in the soft $W$-strahlung-like regions of phase space.}  The reconstructed $Z'$ resonance, after application of all cuts, is shown in figure~\ref{fig:WWresonance}.

\begin{table}
 \centering
\begin{tabular}{|l|c|c|c|}  \hline
 & $m_{Z'} = 1 \ {\rm TeV}$ & $m_{Z'} = 2 \ {\rm TeV}$ & $m_{Z'} = 3 \ {\rm TeV}$ \\ \hline \hline
Signal Eff.   & 12.3\% & 15.8\%  & 15.6\% \\ \hline
$\sigma(Wj:\ qg \to qW)$ & 43.8 fb  & 1.47 fb & 0.11 fb \\
$\sigma(Wj:\ q\bar q \to gW)$ & 13.0 fb & 0.57 fb & 0.050 fb\\
$\sigma(t\bar t)$ & 23 fb & 0.57 fb & 0.011 fb  \\ 
$\sigma(WW) $ & 3.77 fb & 0.33 fb & 0.042 fb  \\
$\sigma(WZ) $ & 1.01 fb & 0.068 fb & 0.0068 fb  \\
$\sigma(Wh) $ & 0.029 fb & 0.0017 fb & 0.00013 fb  \\
\hline
\end{tabular}
\caption{\it Signal efficiency and background cross sections after all cuts in $WW \to (l\nu)(q\bar q')$.  Note that the signal efficiency includes the branching fractions for the $W$ decays.}
\label{tab:lvjjEff}
\end{table}

\begin{table}
 \centering
\begin{tabular}{|c|c|c|c|} \hline
Luminosity & $m_{Z'} = 1 \ {\rm TeV}$ & $m_{Z'} = 2 \ {\rm TeV}$ & $m_{Z'} = 3 \ {\rm TeV}$ \\ \hline \hline
${\cal L} = 30\ {\rm fb}^{-1}$ & 58.3 fb  & 9.02 fb & 2.70 fb  \\
${\cal L} = 100\ {\rm fb}^{-1}$ & 31.9 fb & 4.94 fb & 1.48 fb \\
${\cal L} = 300\ {\rm fb}^{-1}$ & 18.4 fb  & 2.85  fb & 0.85 fb \\
$S/B = 1$ & 501.5 fb &15.44 fb & 1.36 fb \\ \hline 
\end{tabular}
\caption{\it $\sigma(Z') \times BR(Z'\rightarrow W^+W^-)$ required for discovery or $S/B=1$ in the $(l\nu)(q\bar q')$ mode.}
\label{tab:lvjjsigma}
\end{table}

\begin{figure}[tp]
\begin{center}
\epsfig{figure=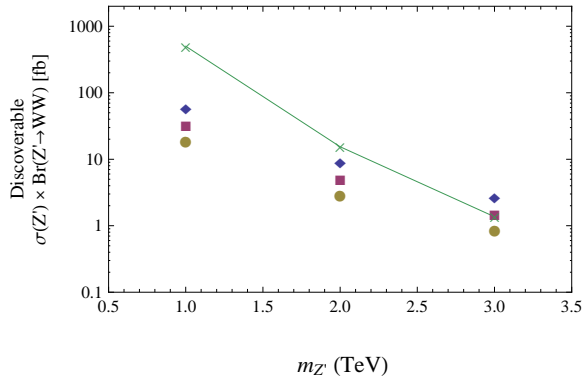,scale=0.60}
\caption{\it Discovery reach for $WW \to (l\nu)(q\bar q')$.  The three symbols (diamond, box, circle) correspond to (30, 100, 300) fb$^{-1}$ of integrated luminosity.  The $\times$ refers to the value that is required to achieve $S/B = 1$.}
\label{dis_reach_lvjj}
\end{center}
\end{figure}

Our signal box is $m_W^{\rm reco} = [65,95]$ GeV and $m_{Z'}^{\rm reco} = [m_{Z'}-15\%,m_{Z'}+15\%]$.  The final signal efficiencies and background cross sections are shown in table~\ref{tab:lvjjEff}.  The discovery reach, i.e.\ the minimum $Z'$ cross section times $BR(W^+W^-)$ required to claim discovery, is listed for integrated luminosities of 30, 100, and 300~fb$^{-1}$ in table~\ref{tab:lvjjsigma}, and displayed graphically in fig~\ref{dis_reach_lvjj}.  We also list/display the $\sigma\times BR$ required to achieve $S/B = 1$.

As one benchmark to interpret the results, we point out that a 3 TeV $Z'$ with $\sigma\times BR \simeq 1.5$~fb, such as the warped 5D case discussed in section~\ref{sec:theory}, can be discovered with about 100~fb$^{-1}$ of data (and with $O(1)$ $S/B$ ratio).  Previous estimates had placed discovery at the ab$^{-1}$ scale~\cite{Agashe:2007ki} or higher~\cite{Benchekroun:2001je}.

%-------------------------------------------------------
\subsection{$Zh$ with leptonic $Z$: $(l^+l^-)(b\bar b)$}
\label{sec:Zhllbb}
%-------------------------------------------------------

\begin{figure}[tp]
\begin{center}
\epsfig{figure=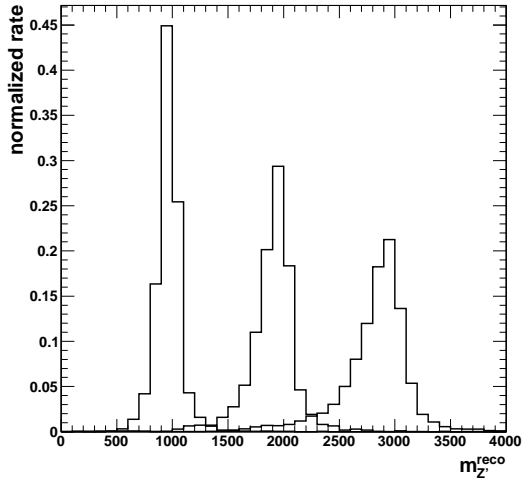,scale=0.37}
\caption{\it Reconstructed $Z'$ invariant masses for 1, 2, and 3 TeV in the $Zh \to (l^+l^-)(b\bar b)$ mode.}
\label{llbb_higgs_zprime}
\end{center}
\end{figure}

For this analysis, we reconstruct the leptonic $Z$ by demanding exactly two isolated leptons with opposite-sign and same-flavor, and which fall into the mass window $m_{l^+l^-} = [80,100]$ GeV.  These should be within $|\Delta\phi| < \pi/2$ of each other, and the leptonic $Z$ reconstructed from them should be $|\Delta\phi| > \pi/2$ away from the Higgs candidate jet.  The $Z'$ is reconstructed as a simple four-vector sum of the leptonic $Z$ with the Higgs-subjets.  As with the \WpWm\ analysis above, we demand that the Higgs candidate satisfy $p_T > m_{Z'}^{\rm reco}/3$.  Figure~\ref{llbb_higgs_zprime} displays the reconstructed resonance peaks after application of cuts.

Searches for a light Higgs decaying to $b\bar b$ traditionally rely heavily on $b$-tagging.  However, the quality of vertex-based $b$-tagging is expected to quickly degrade at high $p_T$, and we do not here have the ability to model this.  Nonetheless, we can ask whether a more robust form of $b$-tagging can be applied, or whether $b$-tagging needs to be applied at all.  In lieu of capitalizing on the displaced vertices of $B$ hadrons, we use the well-exploited fact that they have fairly high semi-leptonic branching fractions, specifically to the very clean case of muons.  A muon can be produced either promptly from the $b$ quark decay, or from a subsequent charm decay, leading to a net $BR$ of about 20\% for $B\to \mu$+anything.  Since the Higgs decay gives us two chances to find one of these decays, our net tag rate for $\ge 1$ muon embedded within the Higgs-subjets should approach 35\%, independent of $p_T$.  Physical backgrounds to this include prompt heavy flavor ($b$- and $c$-jets), gluons splitting to heavy flavor within light QCD jets, and decays-in-flight of pions and kaons (which we include in our {\tt PYTHIA} simulations).  We implement the tag by searching out muons with $p_T > 5$ GeV within cones of size $R = \Delta R_{\rm subjets}$ around each subjet.  In practice, we find tag rates of roughly 30\%, and mistag rates at the level of 3-8\%, increasing slighly with $p_T$ and generally somewhat higher for gluons.  There will also be backgrounds from fake muons, but we assume that these can be brought to a level below the physical mistag rate.

\begin{table}
 \centering
\begin{tabular}{|l|c|c|c|} \hline
& $m_{Z'} = 1 \ {\rm TeV}$ & $m_{Z'} = 2 \ {\rm TeV}$ & $m_{Z'} = 3 \ {\rm TeV}$ \\ \hline \hline
Signal Eff.   & 1.3\% (0.39\%) & 1.7\% (0.53\%) & 1.7\% (0.57\%) \\ \hline
$\sigma(Zj:\ qg \to qZ)$ & 5.68 fb (0.16 fb)  & 0.18 fb (0.0066 fb) & 0.014 fb (0.0006 fb) \\
$\sigma(Zj:\ q\bar q \to gZ)$ & 1.52 fb (0.09 fb) & 0.067 fb (0.0053 fb) & 0.006 fb (0.0005 fb) \\ 
$\sigma(Z(W/Z))$ & 0.093 fb (0.009 fb) & 0.0042 fb (0.00011 fb) & 0.00044 fb (0.000014 fb) \\
$\sigma(Zh) $ & 0.046 fb (0.013 fb) & 0.0024 fb (0.0007 fb) & 0.00025 fb (0.000076 fb)  \\ 
\hline
\end{tabular}
\caption{\it Signal efficiency and background cross sections after all cuts in $Zh \to (l^+l^-)(b\bar b)$.  Note that the signal efficiency includes the branching fractions for the $Z$ and $h$ decays.  The numbers in parentheses are after $\mu$-tagging.}
\label{tab:llbbEff}
\end{table}

\begin{table}
 \centering
\begin{tabular}{|c|c|c|c|} \hline
Luminosity & $m_{Z'} = 1 \ {\rm TeV}$ & $m_{Z'} = 2 \ {\rm TeV}$ & $m_{Z'} = 3 \ {\rm TeV}$ \\ \hline \hline
${\cal L} = 30\ {\rm fb}^{-1}$ & 186.2 fb (121.9 fb) & 27.8 fb (62.4 fb) & 19.7 fb (59.0 fb)  \\
${\cal L} = 100\ {\rm fb}^{-1}$ & 102 fb (66.8 fb)    & 15.2 fb (18.7 fb)  & 5.9 fb (17.7 fb) \\
${\cal L} = 300\ {\rm fb}^{-1}$ & 58.9 fb (38.6 fb)   &   8.8 fb (6.2 fb)     & 2.4 fb (5.9 fb)\\
$S/B = 1$                                & 552.6 fb (69.8 fb)    & 15.3 fb (2.4 fb)    & 1.2 fb (0.2 fb) \\ \hline 
\end{tabular}
\caption{\it $\sigma(Z') \times BR(Z'\rightarrow Zh)$ required for discovery or $S/B=1$ in the $(l^+l^-)(q\bar q')$ mode.  The numbers in parantheses are after $\mu$-tagging.}
\label{tab:llbbsigma}
\end{table}

\begin{figure}[tp]
\begin{center}
\epsfig{figure=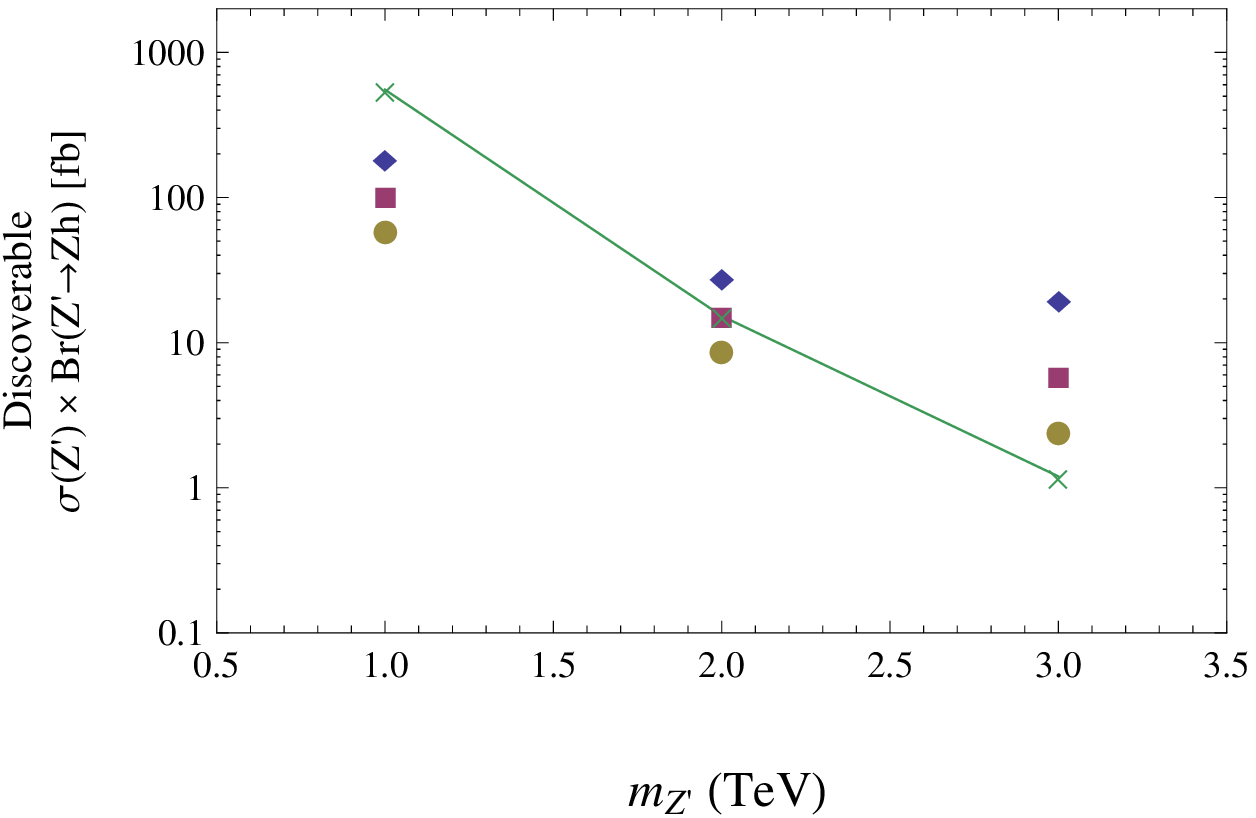,scale=0.60} \hspace{0.5cm}
\epsfig{figure=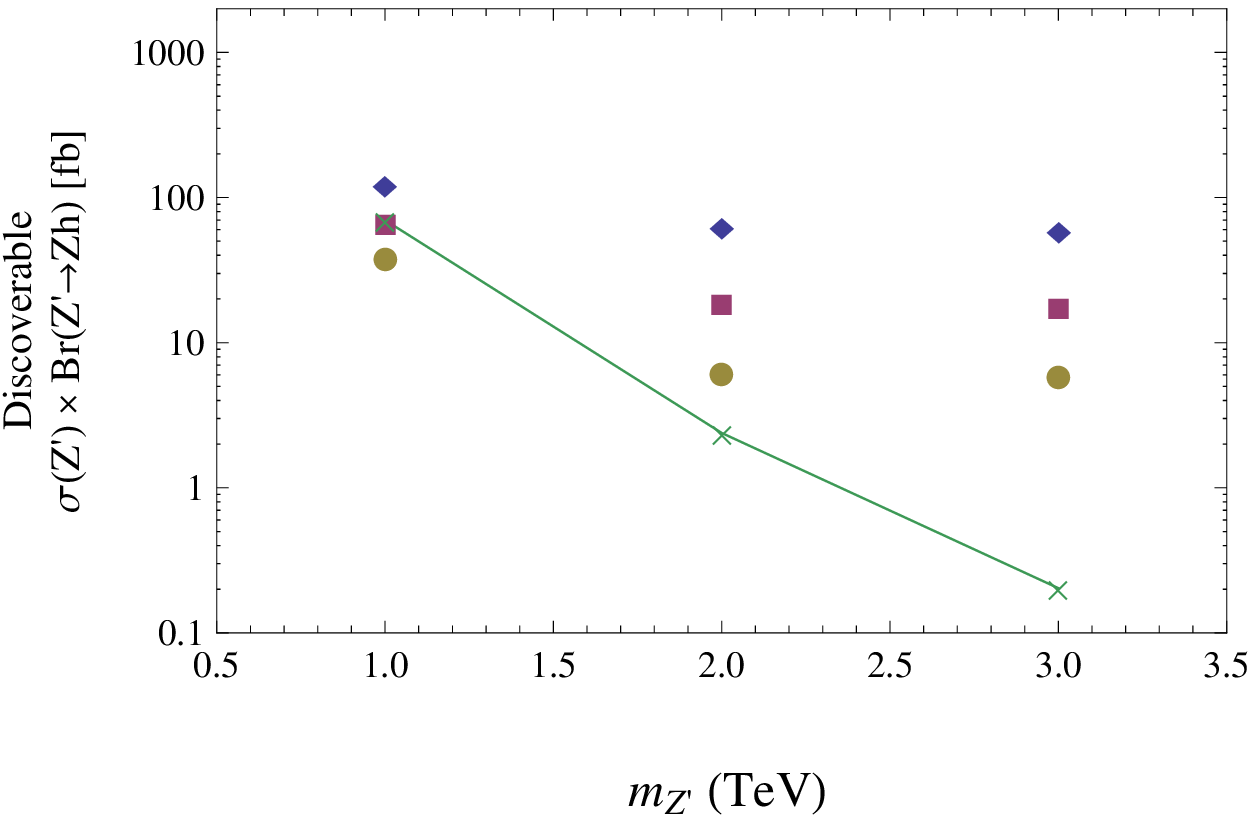,scale=0.60}
\caption{\it Discovery reach for $Zh \to (l^+l^-)(b\bar b)$ without (left) and with (right) $\mu$-tagging.  The three symbols (diamond, box, circle) correspond to (30, 100, 300) fb$^{-1}$ of integrated luminosity.  The $\times$ refers to the value that is required to achieve $S/B = 1$.}
\label{dis_reach_llbb}
\end{center}
\end{figure}

We run two concurrent analyses, one without a $b$-tag and one with the $\mu$-based $b$-tag.  The signal box is $m_h^{\rm reco} = [100,140]$ GeV and $m_{Z'}^{\rm reco} = [m_{Z'}-15\%,m_{Z'}+15\%]$.  The final signal efficiencies and background cross sections are shown in table~\ref{tab:llbbEff}.  The discovery reach is listed in table~\ref{tab:llbbsigma}, and displayed graphically in fig~\ref{dis_reach_llbb}, along with the $\sigma\times BR$ required to achieve $S/B = 1$.

We can see that the statistical reach and $S/B$ are clearly improved by the $\mu$-tag for lower masses, but it becomes overkill at higher masses, where the untagged background is only $O(1)$ events even at 300~fb$^{-1}$.  For 2 and 3 TeV, the tagged discovery reach is essentially controlled by the $N_S = 10$ requirement, with near-vanishing background.  Clearly, $b$-tagging is most important at sub-TeV masses, where the background can still be substantial.  Fortuitously, sub-TeV masses are exactly where traditional $b$-tagging techniques would operate well.  But we learn here that $b$-tagging in this mode at multi-TeV mass may in fact be counterproductive, at least from a strictly statistical standpoint.

Using the same benchmark model as in the previous subsection, 3 TeV and $1.5$~fb$^{-1}$, we find that discovery would require a bit less than 400~fb$^{-1}$.  
This can be contrasted with~\cite{Agashe:2007ki}, which required slightly less than 1~ab$^{-1}$.

%-------------------------------------------------------------
\subsection{$Zh$ with invisible $Z$: $(\nu\bar\nu)(b\bar b)$}
\label{sec:Zhvvbb}
%-------------------------------------------------------------

At low $p_T$, processes involving an invisible $Z$ are challenging because the neutrinos are nearly back-to-back.  In our case, they are nearly collinear, leading to striking monojet signals with a single TeV-scale jet and almost no other activity.  Honestly modeling the background in this case can be somewhat subtle, as non-gaussian fluctuations in energy measurements can create fake missing energy.  However experimental simulation studies (see e.g.~\cite{CMS:monojet}) indicate that generic QCD backgrounds, as well as top and $W$ backgrounds, can be readily controlled.  Consequently, we focus exclusively on $Z$+jets and SM diboson backgrounds.

We use our default analysis cuts, and apply no special cuts to insist on large missing energy.  A more realistic analysis would require some kind of veto on activity opposite the Higgs candidate, but this would still pass the signal with very high efficiency.  We define missing energy as simply the transverse momentum vector that balances the Higgs, $\vec{\displaystyle{\not}E}_T \equiv -\vec{p}_T(h)$.  A more refined measurement would also take into account energy deposited by ISR jets, etc.

\begin{figure}[tp]
\begin{center}
\epsfig{figure=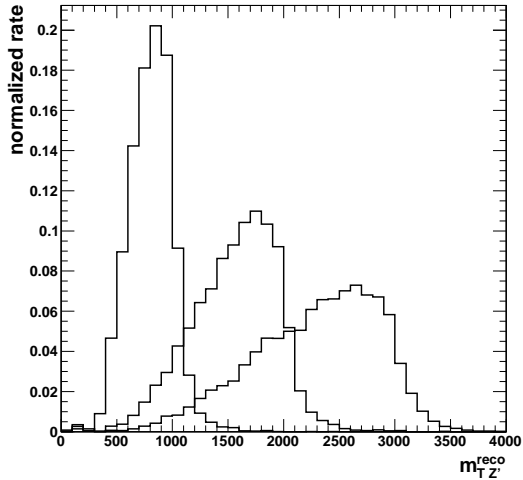,scale=0.37}
\caption{\it Reconstructed $Z'$ transverse masses for 1, 2, and 3 TeV in the $Zh \to (l^+l^-)(b\bar b)$ mode.}
\label{vvbb_comparision_signal}
\end{center}
\end{figure}

Proper reconstruction of the $Z'$ mass peak is impossible, as there is no unambiguous way to know the $p_Z$ of the invisible $Z$.  However, the $p_T$ of the $Z'$ decay products has a Jacobian peak which is spin-enhanced.  In figure~\ref{vvbb_comparision_signal}, we show the distribution of the reconstructed $Z'$ transverse mass,\footnote{We also explored two other types of reconstruction, exploiting the fact that the $Z'$ is produced approximately at rest in the lab frame.  We considered the mass formed by assuming that the $Z$ and $h$ are produced back-to-back in three dimensions in the lab frame, as well as a mixture between this and the transverse mass by using the full $h$ three-vector and the invisible $Z$ transverse vector.  Of our three $Z'$ reconstruction methods, we find that the transverse mass works the best for all of the physical $m_{Z'}$ that we consider.}
\beq
m_{T\,Z'}^{\rm reco} \equiv \sqrt{p_T(h)^2+m(h)^2}+\sqrt{\mathmet^2+m_Z^2} \simeq 2p_T(h).
\eeq

\begin{table}
 \centering
\begin{tabular}{|l|c|c|c|} \hline 
 & $m_{Z'} = 1 \ {\rm TeV}$ & $m_{Z'} = 2 \ {\rm TeV}$ & $m_{Z'} = 3 \ {\rm TeV}$ \\ \hline \hline
Signal Eff.   & 5.4\% (1.7\%) & 6.8\% (2.2\%) &  7.1\% (2.3\%) \\ \hline
$\sigma(Zj:\ qg \to qZ)$ & 43.0 fb (1.54 fb)  & 1.42 fb (0.049 fb) & 0.11 fb (0.0058 fb) \\
$\sigma(Zj:\ q\bar q \to gZ)$ & 9.2 fb (0.46 fb) & 0.38 fb (0.03 fb) & 0.037 fb (0.003 fb) \\ 
$\sigma(Z(W/Z))$  & 0.87 fb (0.044 fb) & 0.064 fb (0.0024 fb) & 0.0037 fb (0.00095 fb)  \\
$\sigma(Zh)$ & 0.073 fb (0.025 fb) & 0.0035 fb (0.00097 fb) & 0.00041 fb (0.0001 fb)  \\ 
\hline
\end{tabular}
\caption{\it Signal efficiency and background cross sections after all cuts in $Zh \to (\nu\bar\nu)(b\bar b)$.  Note that the signal efficiency includes the branching fractions for the $Z$ and $h$ decays.  The numbers in parentheses are after $\mu$-tagging.}
\label{tab:vvbbEff}
\end{table}

\begin{table}
\centering
\begin{tabular}{|c|c|c|c|} \hline
Luminosity & $m_{Z'} = 1 \ {\rm TeV}$ & $m_{Z'} = 2 \ {\rm TeV}$ & $m_{Z'} = 3 \ {\rm TeV}$ \\ \hline \hline
${\cal L} = 30\ {\rm fb}^{-1}$ & 123.7 fb (78.6 fb)  & 18.2 fb (15.5 fb) & 4.97 fb (14.8 fb) \\
${\cal L} = 100\ {\rm fb}^{-1}$ & 67.7 fb (43.1 fb)    & 9.97 fb (6.64 fb)  & 2.7 fb (4.4 fb)\\
${\cal L} = 300\ {\rm fb}^{-1}$ & 39.1 fb (24.9 fb)    & 5.76 fb (3.83 fb)  & 1.57 fb (1.48 fb) \\
$S/B = 1$                                  & 987.7 fb (123.8 fb) & 27.2 fb (3.8 fb) & 2.1 fb (0.43fb) \\ \hline 
\end{tabular}
\caption{\it $\sigma(Z') \times BR(Z'\rightarrow Zh)$ required for discovery or $S/B=1$ in the $(\nu\bar\nu)(b\bar b)$ mode.  The numbers in parentheses are after $\mu$-tagging.}
\label{tab:vvbbsigma}
\end{table}

\begin{figure}[tp]
\begin{center}
\epsfig{figure=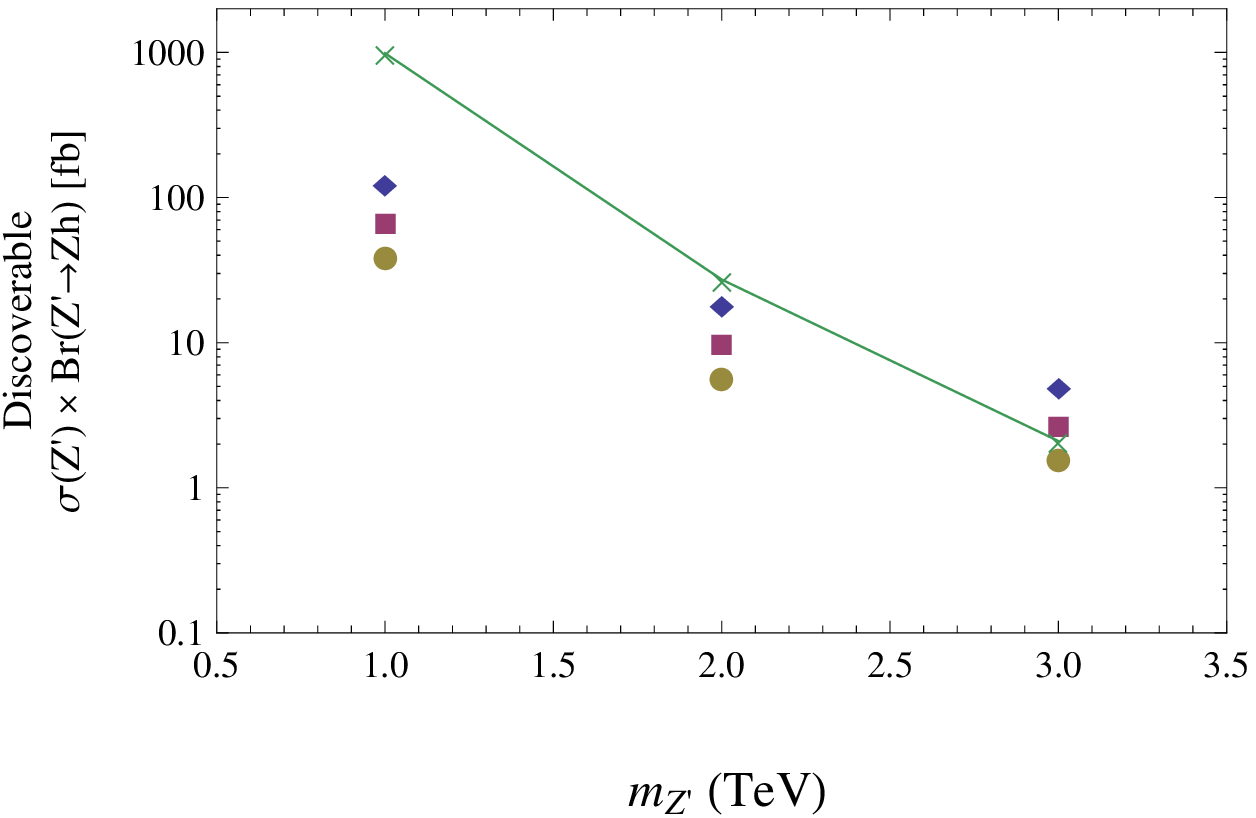,scale=0.60} \hspace{0.5cm}
\epsfig{figure=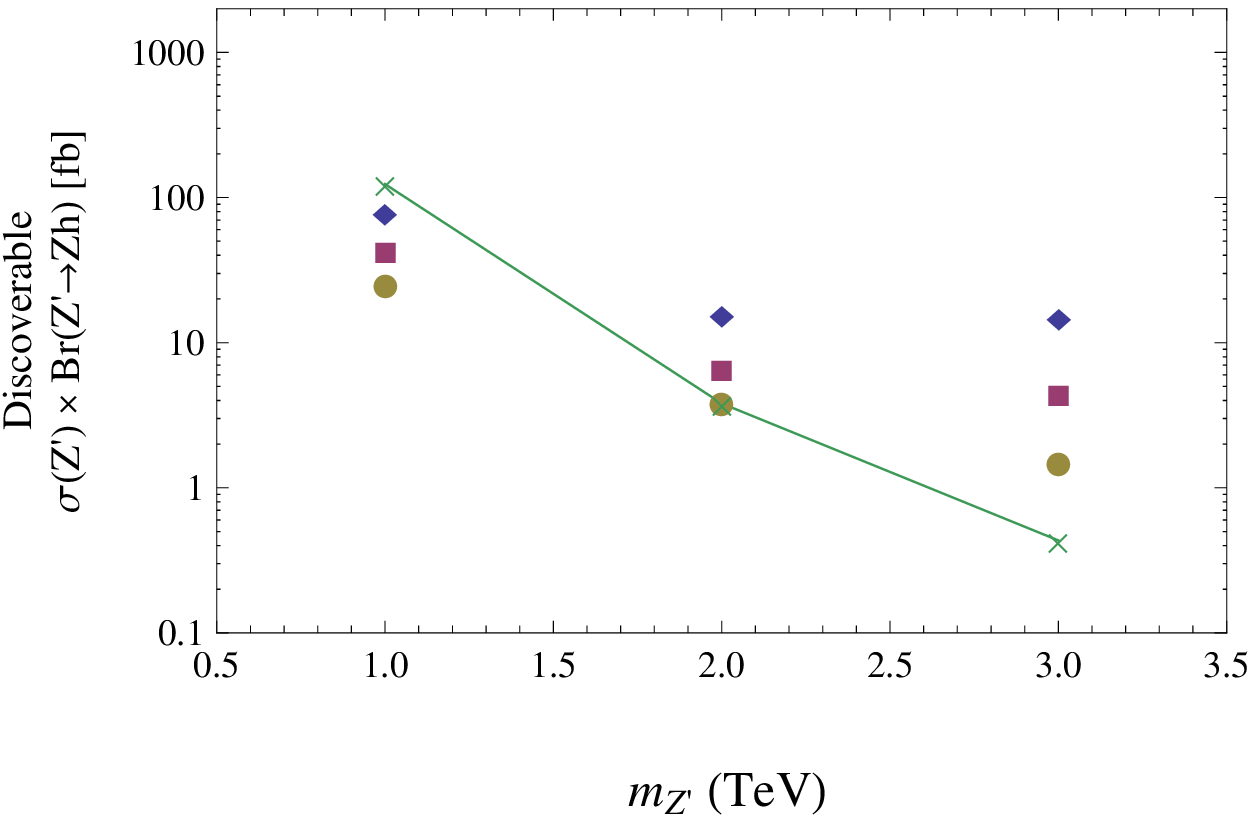,scale=0.60}
\caption{\it Discovery reach for $Zh \to (\nu\bar\nu)(b\bar b)$ without (left) and with (right) $\mu$-tagging.  The three symbols (diamond, box, circle) correspond to (30, 100, 300) fb$^{-1}$ of integrated luminosity.  The $\times$ refers to the value that is required to achieve $S/B = 1$.}
\label{dis_reach_vvbb}
\end{center}
\end{figure}

As a signal box, we take $m_h^{\rm reco} = [100,140]$ GeV and $m_{T\,Z'}^{\rm reco} = [m_{Z'}-30\%,m_{Z'}+10\%]$.  As in the leptonic $Z$ analysis, we consider cases with/without $\mu$-tagging.  The final signal efficiencies and background cross sections are shown in table~\ref{tab:vvbbEff}.  The discovery reach is listed in table~\ref{tab:vvbbsigma}, and displayed graphically in fig~\ref{dis_reach_vvbb}, along with the $\sigma\times BR$ required to achieve $S/B = 1$.

From the perspective of pure statistics, our analysis of $Zh$ with invisible $Z$ suggests that it may be even more sensitive than the corresponding analysis with leptonic $Z$, mainly owing to the three times higher event rate (as well as our lack of reconstruction criteria on the $Z$).  Of course, this comes at a price of lower $S/B$, and less distinctive signal shape, as the kinematic recontruction is less precise.  Still, $\mu$-tagging is no longer as detrimental above 2 TeV, for example allowing discovery of the 3 TeV warped $Z'$ at about 300~fb$^{-1}$ with $S/B \simeq 3$.  We believe that this mode merits a more prominent role for resonance searches at the LHC.

\section{Conclusions}
\label{sec:conclusion}

%%%%%%%%%%%%%%%
% Conclusions 
%%%%%%%%%%%%%%%

In this paper, we have explored the utility of jet substructure methods to aid in the search for a $Z'$ decaying to electroweak bosons at the LHC.  Our findings can be naturally divided into two parts.  First, we have made a detailed categorization of the performance of the BDRS substructure algorithm for identifying boosted hadronically-decaying electroweak bosons in the TeV regime.  Second, we have estimated the $Z'$ discovery potential in several final-state channels, demonstrating significant improvements over previous analyses.

In order to clearly justify the use of substructure, we have made a comparison between BDRS and more traditional jet-based reconstructions.  While the difference is modest for genuine boosted electroweak bosons, we isolated several advantages in the treatment of the background.  For QCD jets with mass near the electroweak boson mass, BDRS achieves $O(1)$ greater discrimination by exploiting the jets' internal kinematics.  Beyond this, since BDRS is free of artificial $\Delta R$ thresholds, the background mass spectrum is free of artificial mass thresholds, and it is also much flatter.

Another important observation is the stability and benefit of using a large fat-jet clustering radius.  We chose a radius of $R=1.4$, which covers an area much larger than a traditional jet, and in particular much larger than a TeV-scale electroweak boson-jet.  In some sense, we are no longer applying ``jet'' substructure but rather approaching ``event'' substructure.  While the entire event is contaminated with plentiful uncorrelated radiation, we nonetheless find that BDRS is extremely efficient at zooming-in to the most interesting region of activity.  The scale-invariant picture of the event offered by this procedure offers several advantages.  As noted above, the background mass spectrum becomes modestly-sloped and largely featureless.  The large catchment area of the fat-jet allows a single analysis strategy to interpolate smoothly between resonance searches in the several-hundred GeV range up to the multi-TeV range.  Finally, a nontrivial amount of global color-discrimination is incorporated for free.

Ultimately, this scale-invariant picture breaks down once we hit the size of the individual detector elements.  We incorporated into our analysis a primitive detector model which highlights this limitation, and suggests how to overcome it.  We used the highly-segmented ECAL as a tracer of spatial energy flow, and incorporated the full ECAL+HCAL energy measurements.  While a procedure like this is not so crucial for a 1 TeV $Z'$, it becomes very important for 3 TeV, where the quarks from the secondary decay of a $W$ will usually sit in adjacent HCAL cells.  We do not expect that our model gives a truly accurate picture of the jet mass resolution at very high $p_T$, but it does demonstrate that the LHC detectors have enough information to resolve the individual quarks.  Our resonance search results should be largely robust if the true mass resolution can come within $O(1)$ of our estimate of roughly 12\%.

For our resonance searches, we explored decays into $Zh$ and \WpWm, the former under the assumption of a 120 GeV SM-like Higgs decaying primarily to $b\bar b$.  In general, we find much better reach than what has previously been claimed.  For example, a 3 TeV $Z'$ with fb-scale cross section need not be relegated to the Super-LHC.  This specific mass has become a kind of benchmark for composite/5D Higgs models, as a minimum to pass electroweak precision tests.   We find that discovery should be possible with as little as 100~fb$^{-1}$ in the \WpWm\ mode, with $O(3)$ times higher luminosity required for $Zh$ to catch up.

Previous studies of $Zh$ have focused exclusively on the $(l^+l^-)(b\bar b)$ final state utilizing $b$-tags.  It is well-known that $b$-tagging performance degrades at high $p_T$, owing to the increasing collinearity of hits in the inner trackers.  In order to bypass the complicated modeling of this behavior, we have focused exclusively on a muon-based $b$-tagger, which should extrapolate robustly to high energies.  We found that this tagger improves discovery reach below 2 TeV, but that it becomes counterproductive at higher masses due to the $O(1/3)$ signal tagging efficiency and the small number of expected untagged background events at normal LHC luminosities.  Of course, it is possible that a more sophisiticated high-$p_T$ $b$-tagger could perform better.

We have also studied the largely neglected mode $(\nu\bar\nu)(b\bar b)$.  By reconstructing the $Z'$ transverse mass, we find that this channel is highly competitive.  The three times higher statistics appears to more than compensate for the less precise kinematics.  In the case that a discovery is claimed of $Z'\to Zh$, visible and invisible $Z$ modes will serve as important cross-checks of each other.

We conclude that jet substructure techniques are very promising for identifying highly boosted electroweak bosons, and that these techniques can be fruitfully applied to $Z'$ searches.  (Though we have not explored the following point in detail, our methods can also be immediately applied to searches for other objects that decay into electroweak bosons, such as $W'$ and KK gravitons, as well as searches in weak boson fusion.)  The scope of our analysis was limited to the case of two-body boson decays into quarks.  This is adequate for $W$ (and $Z$) bosons, but Higgses clearly present a much richer set of possible hadronic and semi-hadronic final states.  Identifying these decays will require a broader set of substructure tools than what we have implemented here.  We relegate their study to future work~\cite{ZprimesTau,Zprimes2}.

\acknowledgments{We are grateful to Kaustubh Agashe, Zackaria Chacko, Zhenyu Han, Hye-sung Lee, Petar Maksimovic, Witold Skiba, and Raman Sundrum for useful discussions.  The authors are also gratetful to the organizers of the workshop ``BOOST2010'' at Oxford.  AK was partially supported by NSF grant PHY-0801323.  MS was supported in part by the DoE under grant No.\ DE-FG-02-92ER40704.  BT was supported by JHU grant No.\ 80020033 and by DoE grant No.\ DE-FG-02-91ER40676.}

\appendix

\section{Pileup and Filtering}
\label{sec:pileup}

%%%%%%%%%%%%%%%%%%%%%%%%%%%%%%%%
% Appendix on pileup/filtering
%%%%%%%%%%%%%%%%%%%%%%%%%%%%%%%%

In the analyses of this paper, we consider $Z'$ masses starting at 1 TeV.  In this kinematic region, we find that the effects of underlying event contamination within the subjets are modest.  However, searches extending to multi-TeV $Z'$ masses will only be made possible by high-luminosity running, where we inevitably face the issue of approximately 20 min-bias pileup collisions superimposed on each $Z'$ candidate event.  Determining the best way to remove the pileup contamination (and the underlying event contamination at lower $m_{Z'}$) is beyond the scope of this paper.  However, we do pause here to consider how large the effect is and how effective filtering is at correcting it.

We simulate pileup by adding particles from {\tt PYTHIA} min-bias events to our $Z'$ events.  The number of min-bias collisions sampled is drawn, on an event-by-event basis, from a Poisson distribution with mean of 20.  In principle, charged pileup can be tracked back to event vertices distinct from the $Z'$ production and subtracted, leaving over an irreducible neutral pileup component.  On the other hand, the detectors themselves perform part of this subtraction ``for free,'' by sweeping away soft charged particles in the magnetic field.  To coarsely model this, we also consider events where all charged particles below $p_T = 1$ GeV, including those from the primary hard collision, are removed by-hand prior to jet clustering.\footnote{The magnetic field and inner HCAL radius of the ATLAS and CMS experiments are, respectively, 2.3m/2T and 1.8m/4T.  With these parameters, charged hadrons with $p_T < 0.7$ GeV and $p_T < 1.1$ GeV will not reach the HCAL.}  To factorize this discussion from the possible degrading effect of the magnetic field on the mass reconstruction (as well as all other detector-related issues, which we discuss in appendix~\ref{sec:detector}), we work at particle-level without altering the particles' trajectories.

\begin{figure}[tp]
\begin{center}
\epsfxsize=0.44\textwidth\epsfbox{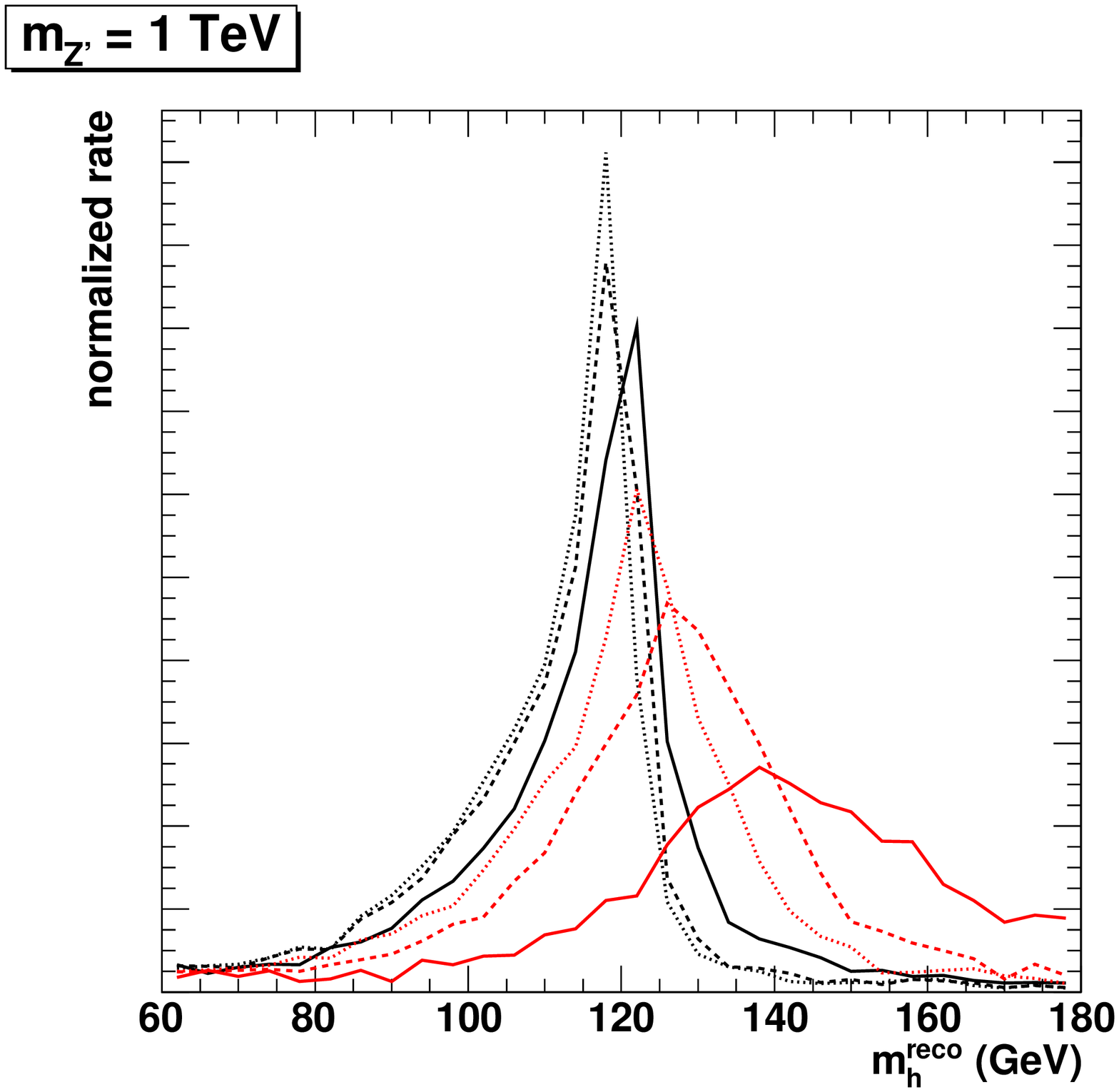}
\epsfxsize=0.44\textwidth\epsfbox{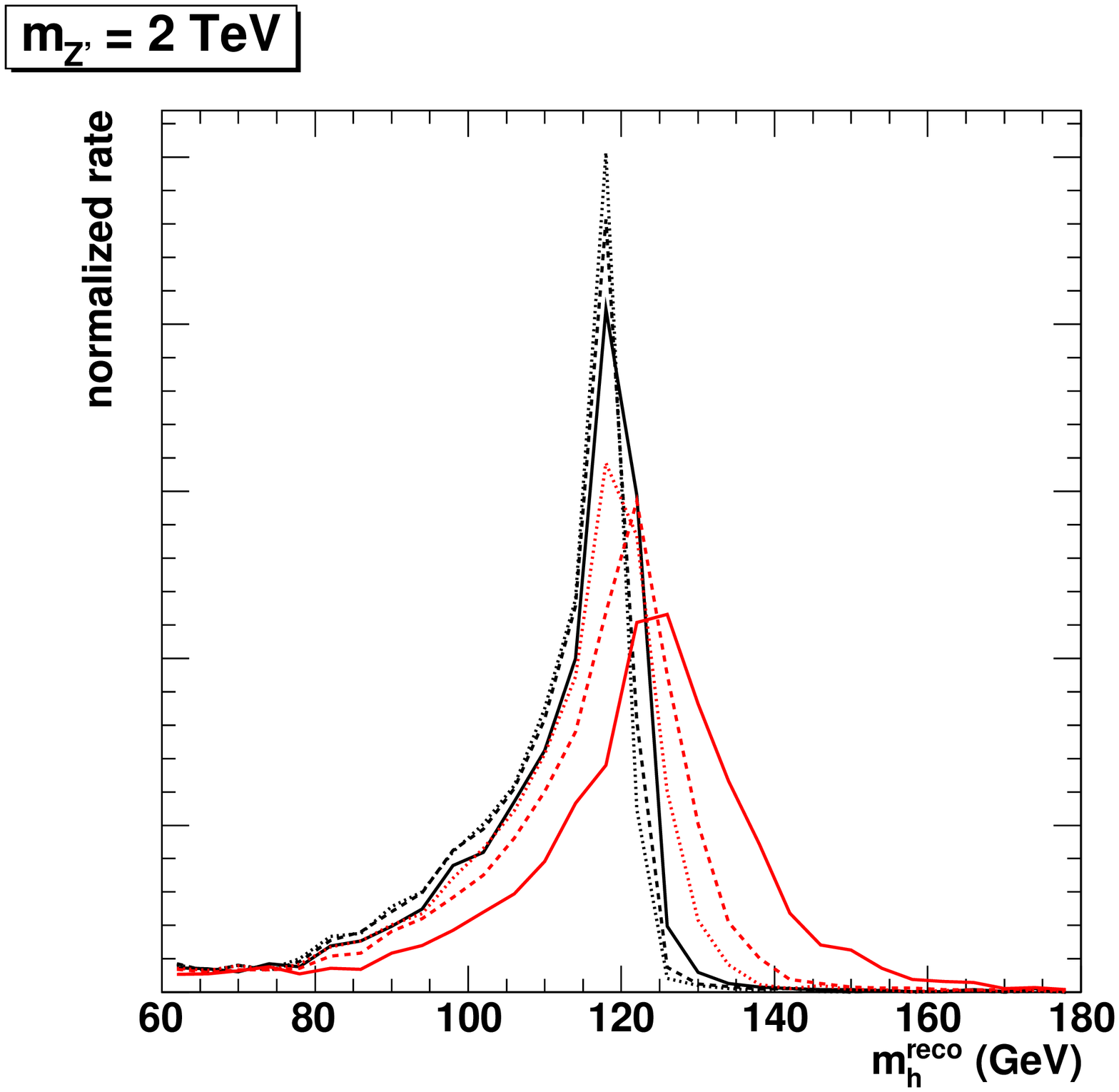}
\caption{\it Distributions of the (particle-level) reconstructed Higgs mass for 1 and 2 TeV $Z'\to Zh \to (l^+l^-)(b\bar b)$.  Displayed are results without pileup (black) and with pileup (red).   Results are further subdivided into:  before filtering (solid), after filtering (dashed), and after filtering with a soft track cutoff (dotted).}
\label{fig:pileup}
\end{center}
\end{figure}

We study the effect of pileup on the reconstructed Higgs mass in the 1 and 2 TeV $Z' \to Zh \to (l^+l^-)(b\bar b)$ samples.  (The $WW$ mode is less sensitive to this contamination at a given $Z'$ mass because $\Delta R_{\rm subjets}$ tends to be smaller.  Partially this is because the $W$ is lighter than the Higgs, but also because the $W$ is highly longitudinally polarized.)  The results are presented in Fig.~\ref{fig:pileup}.\footnote{Note that the uncontaminated mass peak already displays an asymmetric broadening.  This is due to physics, not a flaw in the reconstruction.  Most Higgs decays through $b\bar b$ contain at least one neutrino in the final state.  Some of the broadening is also due to underlying event contamination, particularly at 1 TeV.  Filtering partially corrects this off.}  The effects of pileup are, unsurprisingly, most pronounced in the 1 TeV sample.  The Higgs mass peak becomes shifted upward by about 20 GeV, and broadened by a comparable amount.  The effect is roughly cut in half by filtering, and cut in half again by applying the soft track cutoff.  This reduces the shifting/broadening down to a level that is very likely below that of the realistic experimental resolution.  We have also investigated the case with purely neutral pileup (not shown in Fig.~\ref{fig:pileup}), and find that it looks quite similar to the track-cutoff case.

\begin{figure}[tp]
\begin{center}
\epsfxsize=0.44\textwidth\epsfbox{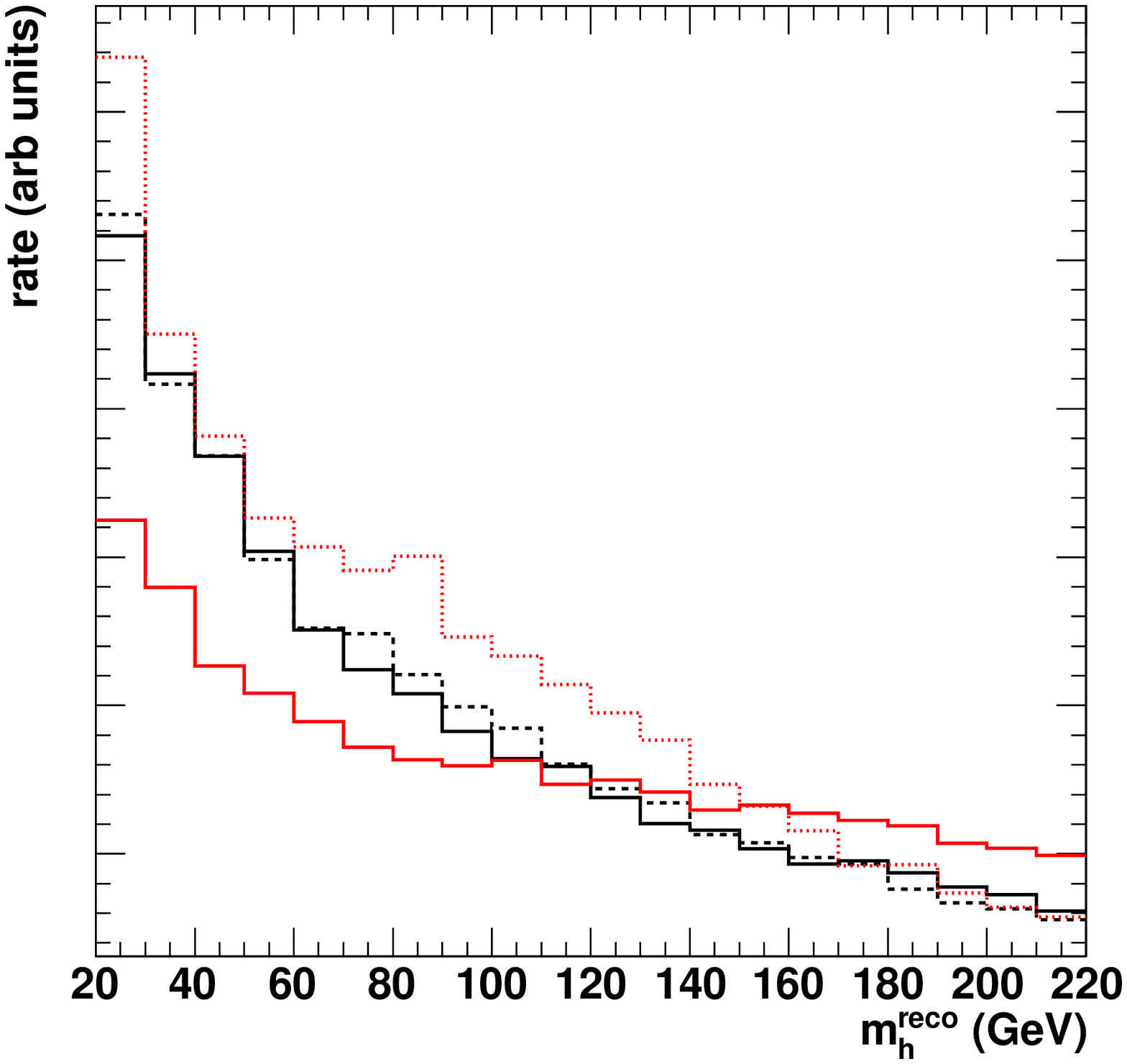}
\epsfxsize=0.44\textwidth\epsfbox{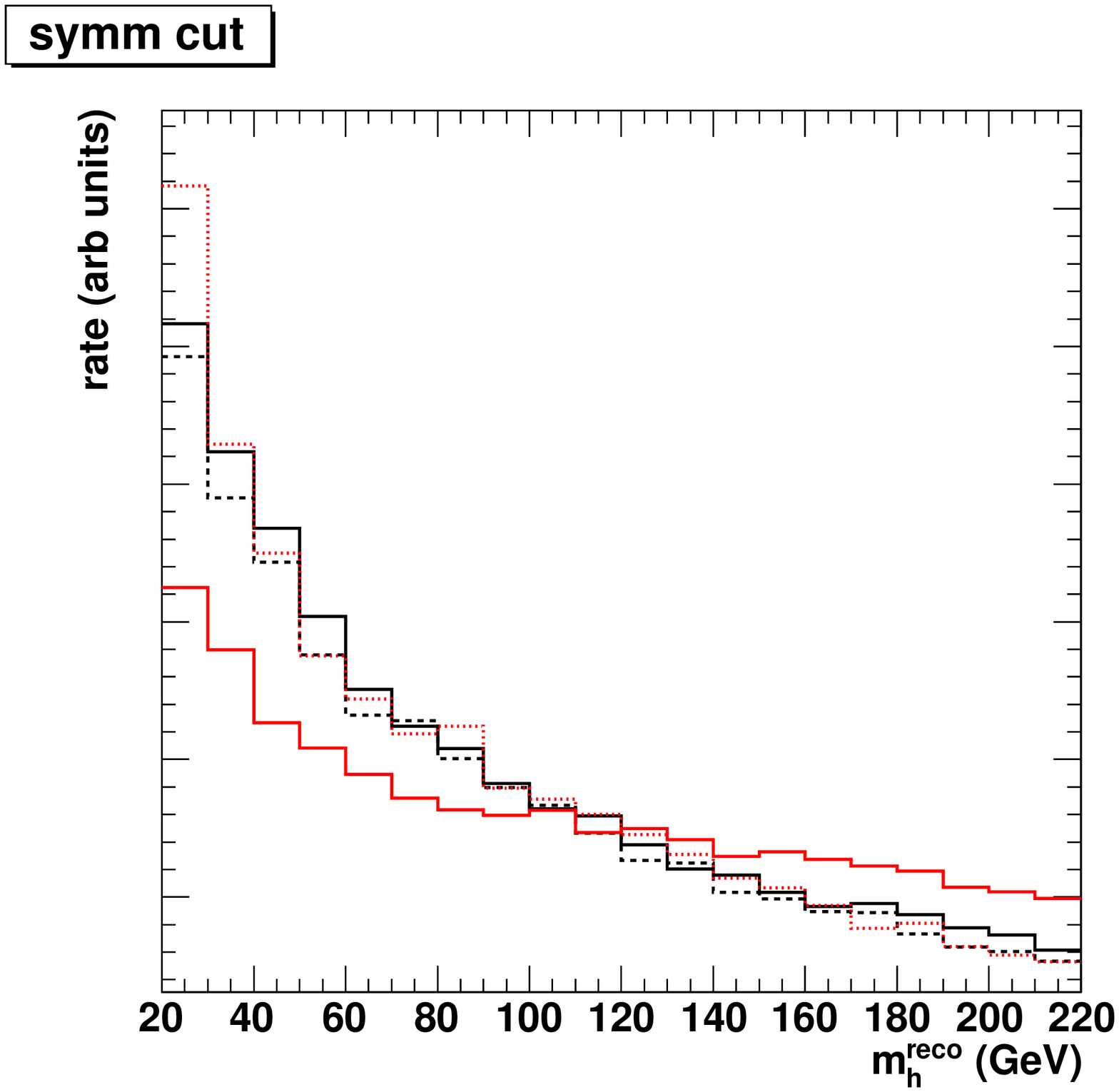}
\caption{\it Distributions of the (particle-level) reconstructed Higgs mass for $Z$+jets $\to(l^+l^-)$+jets backgrounds generated at $\sqrt{\hat s} = 1$ TeV, without (left) and with (right) a symmetry cut applied after filtering.  Displayed are results without pileup (black) and with pileup (red).   Results are further subdivided into:  before filtering (solid), after filtering (dashed), and after filtering with a soft track cutoff (dotted).  (Omitted for clarity are no-pilup with filtering and track cutoff (black-dotted), and pileup with filtering and without track cutoff (red-dashed).)  Note that the arbitrarily-normalized vertical scales on the two plots are identical.}
\label{fig:pileup2}
\end{center}
\end{figure}

The effect of pileup on background reconstruction is more subtle.  QCD subjet configurations with high mass prefer to be highly asymmetric in energy, and the most extreme cases are automatically ignored by the BDRS procedure.  However, such configurations can be pushed above the symmetry threshold $y_{\rm cut}$ by the addition of pileup particles to the softer subjet.  This can be particularly problematic with our very large fat-jet size ($R = 1.4$).  Filtering, even combined with a low-$p_T$ track cutoff, is not powerful enough to fully counteract this.  As an example, we show in the left panel of Fig.~\ref{fig:pileup2} the $m_h^{\rm reco}$ spectrum from our $Z$+jet background generated near $\sqrt{\hat s} = 1$ TeV.\footnote{For the purpose of this study, these events are processed without $b$-tagging and without isolation requirements on the leptons, the latter in order to factorize out changes in lepton identification efficiency in the presence of pileup.  We pick a fixed slice of parton-level $\sqrt{\hat s}$ instead of a fixed slice of $m_{Z'}^{\rm reco}$ in order to disentangle biases introduced by pileup on the latter quantity.  (Such bias is largely corrected away by filtering, but this is difficult to see simultaneously without a full two-dimensional plot.)  While we do not show results at $\sqrt{\hat s}$ above 1 TeV, we indeed find that the effects of pileup become milder, as we did for the $Z'$ samples.}   After application of the track cutoff and filtering, there is a leftover enhancement of the background spectrum near 100 GeV.  We find that the excess is dominated by the cases described above, but that these can easily be removed by insisting that the second-hardest subjet remain fairly hard after filtering.  For example, imposing a post-filtering symmetry cut $p_T($filtered subjet \#2$) / p_T($filtered jet$) > 0.1$ is adequate, as shown in the right panel of Fig.~\ref{fig:pileup2}.  Without pileup, this cut is largely redundant with $y_{\rm cut}$ and has small effect.  The effect on the $Z'$ signal with pileup is at the 1\% level.

Having demonstrated that the degrading effects of pileup can probably be mostly removed for TeV-scale $Z'$ masses, we do not incorporate pileup or filtering into our boosted electroweak boson studies.  However, we do emphasize that any realistic broad-spectrum $Z'$ search should ideally utilize a pileup/UE subtraction method that works well in the sub-TeV mass range but applies smoothly up to trans-TeV masses without adversely affecting the reconstruction there.  We also note that the alternative procedures of pruning~\cite{Ellis:2009su} and trimming~\cite{Krohn:2009th} are also worth exploring, possibly in combination with filtering and with each other~\cite{Soper:2010xk}.

\section{Detector Effects}
\label{sec:detector}

%%%%%%%%%%%%%%%%%%%%%%%%%%%%%%
% Appendix on detector model 
%%%%%%%%%%%%%%%%%%%%%%%%%%%%%%

\begin{figure}[tp]
\begin{center}
\epsfxsize=0.44\textwidth\epsfbox{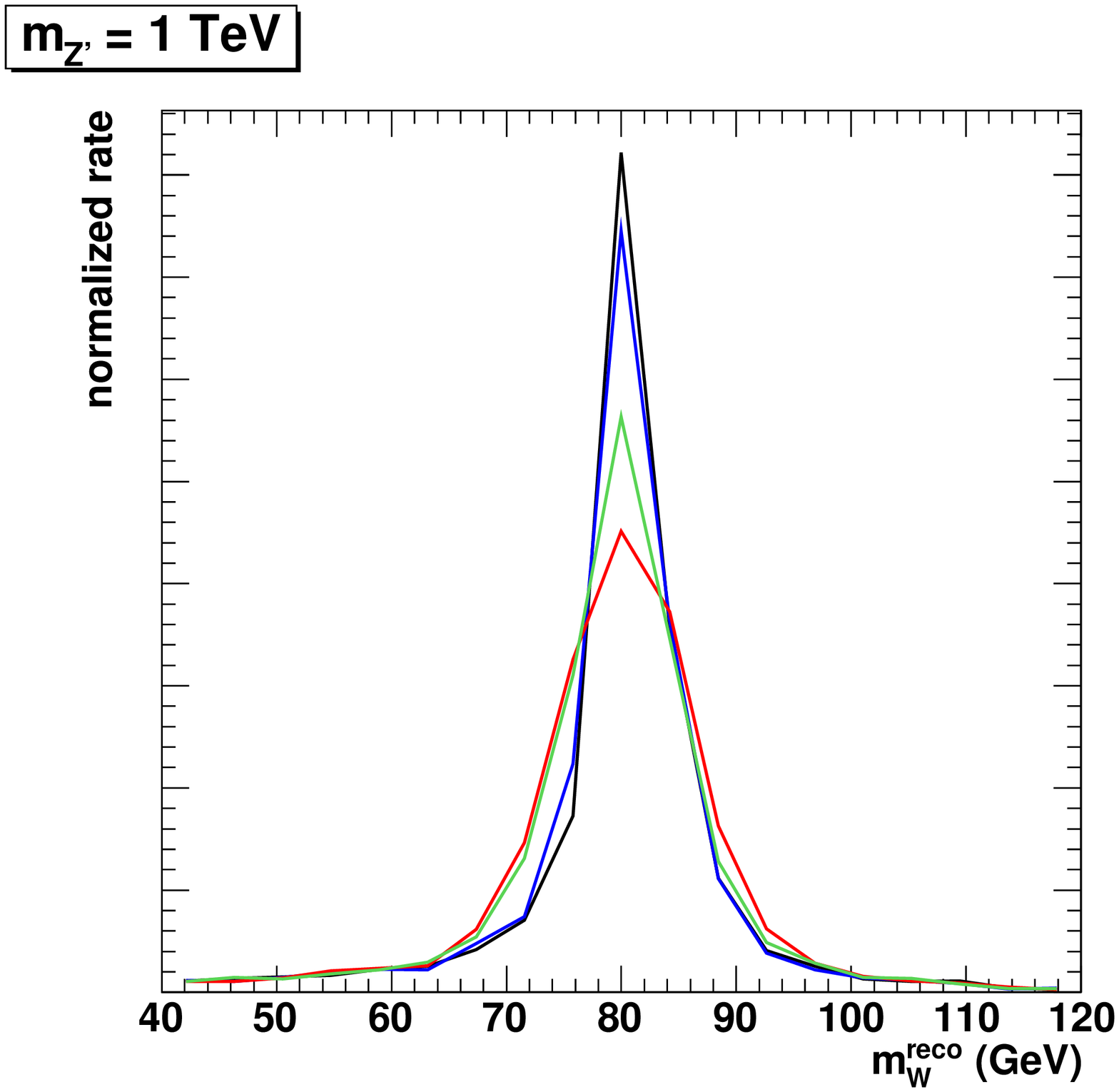}
\epsfxsize=0.44\textwidth\epsfbox{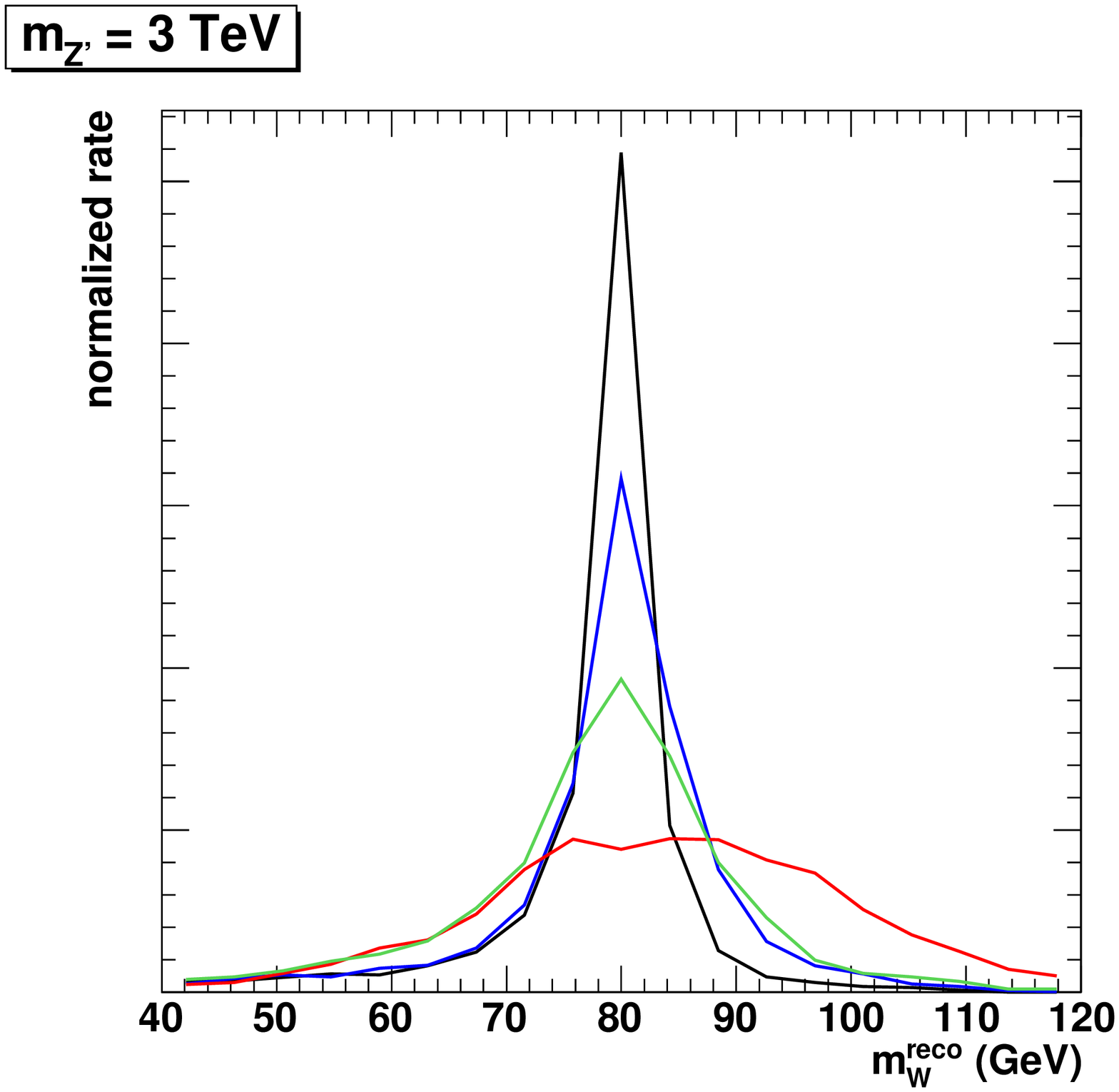}
\caption{\it Distributions of the reconstructed hadronic $W$ mass for 1 and 3 TeV $Z'\to WW \to (l\nu)(q\bar{q}')$.  Displayed are particle-level (black), idealized particle-flow (blue), rescaled ECAL (green), and pure HCAL (red).  Detector models are described in more detail in the text.}
\label{fig:detector}
\end{center}
\end{figure}

While we do not have the tools to fully address the impact of the detector on the quality of boosted boson reconstruction, we can at least make plausibility arguments and highlight what issues might affect the jet-mass measurement.  Probably the biggest worry is that for TeV-scale $W$-jets, the spatial resolution of the detector will be too poor to resolve the mass peak.  To get a feeling for how bad the situation might be, we consider four toy models (inspired by the CMS detector) with decreasing spatial resolution:
\begin{enumerate}
\item  Simple particle-level as a baseline.  
\item  An idealization of particle-flow:  charged particles are measured perfectly, photon energy is deposited into perfect $\Delta\eta\times\Delta\phi = 0.02\times0.02$ ECAL cells, and neutral hadron energy is deposited into perfect $0.1\times0.1$ HCAL cells.  The calorimeter cells then serve as massless particles (with momentum vector oriented along the cell's center in $\eta$-$\phi$) for the purposes of jet clustering.  
\item  An idealized ECAL+HCAL model where ECAL cells are used to trace the energy flow.  Photons and non-isolated electrons are deposited into the ECAL, and all hadrons are deposited into the HCAL.  (Muons and isolated electrons are kept as tracks.)  Each HCAL cell has an associated $5\times5$ block of ECAL cells.  The ECAL energy is rescaled to match the ECAL+HCAL energy, and the HCAL cells are discarded.  The rescaled ECAL cells are used in jet clustering.
\item  An idealized pure HCAL, where all particles (except for muons and isolated electrons) are deposited in $0.1\times0.1$ cells.
\end{enumerate}
We apply these four models to the 1 and 3 TeV $Z'\to WW \to (l\nu)(q\bar{q}')$ signal samples, as shown in Fig.~\ref{fig:detector}.  As one would expect, the impact of the detector model becomes greater at higher $Z'$ mass.  For 1 TeV $Z'$, the typical $\Delta R_{\rm subjets}$ is about 0.35, and our detector segmentation is not a major issue.  For 3 TeV $Z'$, the typical $\Delta R_{\rm subjets}$ is about 0.12, barely large enough for the two subjets to sit in individual HCAL cells.  Nonetheless, even the pure HCAL detector manages to reconstruct the $W$ peak, with about 15 GeV resolution.  The simple trick of using the ECAL to trace the energy flow improves resolution by nearly a factor of 2, down to 8 GeV, and idealized particle flow incorporating the tracker could in principle reduce the smearing to the 5 GeV level.  In what follows, we will take the rescaled ECAL detector as our working model, as a compromise.

Of course, even this model, which neglects the possible utilization of the tracker, is likely optimistic.  This is mainly for three reasons:  the magnetic field will to some extent spread out the charged particles, calorimeter energy deposits will spread laterally as the showers develop inside the cells, and energy measurements in individual cells are subject to sampling errors and electronics noise.  We address each of these in turn.

To coarsely model the effect of the magnetic field, we rotate the momentum vectors of charged particles in $\phi$ so that they are oriented toward the point at which they would impact the innermost part of the calorimeter on an arced trajectory.  We take an inner calorimeter radius of 2 meters, and a magnetic field strength of 4T.  Particles of $p_T < 1.2$ GeV spiral-out and are lost.  Using the rescaled ECAL model and 1 TeV $Z'$, the reconstructed $W$ mass is shifted upward by about 3 GeV and broadened by a comparable amount.  As this is a small effect, and can probably be largely corrected off using particle flow techniques,\footnote{The largest degrading effect is from the softer particles, which separate out from the core of the jet.  These are less susceptible to crowding of hits, which can frustrate precision tracking.} we continue to model the detector without a magnetic field in what follows.

The energy deposited in the calorimeter from a given particle is not usually contained in a single cell, but will leak to some extent into neighboring cells.  To get a sense for how sensitive we might be to this leakage, we rerun the 3 TeV $Z'$ sample through the rescaled ECAL model with ECAL and HCAL cells both doubled in lateral size.  The net effect is a roughly 50\% increase in the $m_W^{\rm reco}$ smearing, to about 12 GeV.  This is comparable to the pure $0.1\times0.1$ HCAL model, and is still very reasonable.\footnote{A further concern is that a very energetic QCD background jet will appear artificially massive due to its energy spreading into several cells.  For example, a very tight 1 TeV jet that shares its energy equally between two adjacent $0.1\times0.1$ HCAL cells would appear to have a mass of 50 GeV, even if its true mass is much smaller.  Incorporating the energy pattern in the ECAL then becomes crucial.  As long as this is done, we do not expect that spatial smearing effects will have such a significant impact on the background jet-mass distribution.}

Finally, we also consider the the effects of smearing out the energy measurements.  The energy resolution of subjets has not been well-studied in full simulation, so as a rough guess we simply use the resolution curve for $R=0.5$ cone jets in the CMS detector~\cite{CMSTDR}:  $\Delta E/E = 5.6/E \oplus 1.25/\sqrt{E} \oplus 0.033$, with $\oplus$ indicating quadrature sum and $E$ measured in GeV.  The effect on the rescaled ECAL results at 3 TeV is modest, whereas the relative effects at smaller $Z'$ mass are more pronounced.  The net result is that the $W$ mass resolution comes out to about 10 GeV (12\%) for all of our $Z'$ samples.  The spatial segmentation effects and energy-sampling effects roughly compensate for each other.

Our nominal choice of detector model for the $Z'$ search is the rescaled ECAL model with energy smearing applied as above.  This roughly approximates some of the dominant resolution effects.  We note that an ATLAS substructure-based study of a $1.1$ TeV $WW$ resonance in vector boson fusion~\cite{Aad:2009wy} quotes a resolution for $m_W^{\rm reco}$ of $7.4$ GeV, which is better than what we obtain with our model.  However, a full simulation study at higher resonance mass is clearly warrented.

%-------------------------

%Acknowledge the UW and UK conferences.

%%%%%%%%%%%%%%
% References
%%%%%%%%%%%%%%

\bibliography{lit}
\bibliographystyle{apsper}

\end{document}